 \documentclass{aa}

\usepackage{graphicx}
\usepackage{amssymb}
\usepackage{txfonts}
\usepackage[round]{natbib}
\bibpunct{(}{)}{;}{a}{}{,}    
\newcommand{\about}{$\simeq$}
\def\degre{^\circ}
\def\degree{\relax\ifmmode^\circ \else $^\circ$\fi}
\newcommand{\Msol}{M$_{\odot}$\ }
\newcommand{\Ti}{$^{44}$Ti\ }

\begin{document}

\title{Are \Ti-Producing Supernovae Exceptional?}

\author{L.-S. The \inst{1}  
\and D.D. Clayton\inst{1} \and R. Diehl\inst{2} 
\and D.H. Hartmann\inst{1} \and A.F. Iyudin\inst{2,3} 
\and M.D. Leising\inst{1}  
\and B.S. Meyer\inst{1} 
\and Y. Motizuki\inst{4} 
\and  V. Sch\"onfelder\inst{2}
}
\institute{Department of Physics and Astronomy, 
 Clemson University, Clemson, SC 29634-0978, USA	
 \and 
 Max-Planck-Institut f\"ur extraterrestrische Physik,
              Postfach 1312, D-85741 Garching, Germany 
 \and
Skobeltsyn Institute of Nuclear Physics, Moscow State University,
Vorob'evy Gory, 119992 Moscow, Russia
 \and 
Cyclotron Center, RIKEN, Hirosawa 2-1, Wako 351-0198, Japan 
}

\offprints{L.-S. The}
\mail{tlihsin@clemson.edu}
\date{Received December 1, accepted December 15, 2005}
\authorrunning{The et al.}
\titlerunning{$^{44}$Ti Supernova Remnants}
\abstract { 
According to standard models supernovae produce radioactive $^{44}$Ti, 
which should be visible in gamma-rays following decay to $^{44}$Ca 
for a few centuries.
\Ti production is believed to be the source of cosmic $^{44}$Ca, whose
abundance is well established. 
Yet, gamma-ray telescopes have not seen the 
expected young remnants of core collapse events. The \Ti mean life of 
$\tau$\about~89~y and the Galactic supernova rate of \about~3/100~y imply 
\about~several detectable \Ti gamma-ray sources, but only one is clearly 
seen, the 340-year-old Cas A SNR. 
Furthermore, supernovae which produce much \Ti are expected to occur primarily 
in the inner part of the Galaxy, where young massive stars are 
most abundant. Because the Galaxy is transparent to gamma-rays, this 
should be the dominant location of expected gamma-ray sources. 
Yet the Cas A SNR as the only one source is located far from 
the inner Galaxy (at longitude 112\degree).
We evaluate the surprising absence of detectable supernovae from
the past three centuries.
We discuss whether our understanding of SN explosions, their
\Ti yields, their spatial distributions, and statistical arguments
can be stretched so that this apparent disagreement may be accommodated within
reasonable expectations, or if we have to revise 
some or all of the above aspects to bring expectations in agreement with the 
observations.
We conclude that either core collapse supernovae have been improbably 
rare in the Galaxy during the past few centuries, or \Ti-producing supernovae 
are atypical supernovae.
We also present a new argument based on $^{44}$Ca/$^{40}$Ca ratios
in mainstream SiC stardust grains that may cast doubt on 
massive-He-cap Type I supernovae as the source of most galactic $^{44}$Ca.
}

\maketitle

\section{Introduction}
\label{sec:intro}
   
   Supernovae are the agents that drive the evolution of gaseous 
   regions of galaxies. As end points of the evolution of massive stars 
   that have formed out of the interstellar gas, their explosions eject 
   matter enriched with freshly formed isotopes and stir 
   interstellar gas. 
   However, the explosions themselves are still not 
   understood \citep{2000Natur.403..727B, 2003fthp.conf...39J}. 
   Parametric descriptions are used to describe the core collapses 
   (``cc-SN'', supernovae 
   of types II and Ib/c) \citep{1995ApJS..101..181W, 1996ApJ...460..408T}
   as well as thermonuclear explosions of white dwarfs (supernovae of type Ia;
   \citealt{1997NuPhA.621..467N}).
   A prominent issue in astrophysics is whether the supernova explosion itself 
   is a well-regulated, robust physical process, 
   or if intrinsic variability over a 
   wider range of physical conditions are rather common. 
   
   Supernova homogeneity by type can be studied in different ways. 
   One approach is to analyze the rate of supernovae of a specific type in 
   different environments and over different time scales.
   In this work we do this by asking 
   if the current rate of supernovae in our Galaxy
   which produce radioactive \Ti is in line with expectations from other 
   observables and from supernova theory. 
   
   \Ti decay offers a unique window to the study of supernova rates.  
   Specific aspects of this window are:
   \begin{enumerate}
   \item Gamma rays penetrate the entire galactic disk with little
         extinction 
   \item \Ti gamma-rays reflect the current rate of supernovae, with the
         \Ti mean decay time scale of $\tau$ = 89 years; 
         this present-day snapshot 
	 which has not yet fed back into chemical evolution can 
         be directly related to the observable current population 
         of massive stars 
   \item Most \Ti is co-produced with $^{56}$Ni in relatively frequent core 
         collapse supernovae (cc-SN). The radioactive energy of 
         $^{56}$Ni is responsible 
	 for well-observed supernova light 
   \item Nucleosynthesis of \Ti is primarily 
	 from $\alpha$-rich freeze-out of nuclear statistical equilibrium and 
         secondarily from silicon burning
   \item The origin of abundant cosmic $^{44}$Ca occurs mainly through \Ti 
         nucleosynthesis
   \item \Ti traces have been found in pre-solar grains which have been 
         attributed to condensation within core-collapse supernovae
   \end{enumerate}

   In this paper we estimate what the gamma-ray sky of \Ti sources 
   would be expected to look like by adopting an {\it average \Ti source model}
   having a characteristic source event 
   {\it recurrence rate, \Ti yield per event}, 
   and {\it spatial distribution}. We compare this to the
   present-day gamma-ray survey and find apparent and serious conflicts. 
   Then we analyze whether deviations from these {\it average} 
   expectations can occur from the known or expected 
   {\it variability of models and 
   parameters involved}. We use a Monte Carlo simulation of the expected
   sky image within reasonable distributions of parameters for that purpose.
   This leads us to discuss each of the relevant parameters, which may explain
   an anomalous  \Ti sky; these are, specifically:
   \begin{itemize}
   \item the gamma-ray survey quality
   \item statistical effects of small samples
   \item the adopted supernova rates
   \item the supernova explosion and nucleosynthesis models
   \item the spatial distribution of supernova events
   \item the deviations from smooth chemical evolution 
   \item the supernova origin of Galactic and solar $^{44}$Ca
   \end{itemize}

\section{\Ti and Supernovae in the Galaxy}
\label{sec:ticonflict}
   
Supernova rates in the Galaxy can be inferred from different observables.
But observational incompleteness and bias requires that several assumptions
are an essential part of such inferences.  
We compare here the observed \Ti sky with expectations for the occurrences
of young \Ti emitting supernova remnants, as they result from astrophysical
assessments of supernova characteristics for our Galaxy. 
  We choose a time interval unit of 100 years for comparison of chemical
  history with supernova event rates.
  The extrapolation of the past history and yields of \Ti-ejecting
  supernova events should give a production rate which can be compared
  with recent supernova rates, as no other
  source of $^{44}$Ca has yet been identified. 

  Have the $^{44}$Ca-producing supernovae been typical;
  is all of $^{44}$Ca produced through radioactive $^{44}$Ti;
  or do exceptional events contribute most of the \Ti?
  Gamma-ray surveys for \Ti sources, and presolar supernova grains provide
  ways to address this question.

In the following Section we will then examine each of the critical assumptions in more detail.
 
For a \Ti sky reference, we adopt the result from COMPTEL's survey 
in the 1.157 MeV band, which is the most complete survey to date 
\citep{1997A&A...324..683D, 1999ApL&C..38..383I}.
In this survey, one object has been clearly detected (340-year-old Cas~A at 
a distance of 3.4~kpc), candidates at lower significance have been discussed
(most prominently GRO~J0852-4642 in the Vela region, 
\citet{1999ApL&C..38..383I}), 
and a weak signal from the Per OB2 association
\citep{1997A&A...324..683D}.
Apparently, no bright young \Ti emitting supernova remnants are found in the
inner region of the Galaxy (see Fig. \ref{fig:timap_IyudinRingberg}).

What do we expect the Galaxy to look like in \Ti emission?

\subsection{\Ti from Supernovae} 
\label{sec:Tiproduction}

{\it Supernova observations} directly demonstrate
   that these events are at the origin of \Ti production:
   The 1.157 MeV $\gamma$-ray line following 
   \Ti decay has been detected in the 340-year old Galactic supernova remnant 
   Cas A \citep{1994A&A...284L...1I, 2001ApJ...560L..79V}.
   Furthermore, SN1987A's late light curve, observed in unique detail 
   over more than 15~years, appears powered by
   a similar amount of $^{44}$Ti (0.2--2.0~10$^{-4}$M$_{\odot}$), 
   from modeling of
   radioactive energy deposition and photon transport 
   in the SNR \citep{1989ApJ...346..395W, 2002NewAR..46..487F}.
   Gamma-ray detection and proof of this interpretation 
   is still lacking, but INTEGRAL's
   recent observations may prove sufficiently sensitive. 
   
{\it Presolar grains} have been identified in meteoritic samples through
     their unusual isotopic abundance patterns, 
     and hold rich isotopic abundance
  detail for characterizing their condensation environments 
  \citep[e.g.][]{2004ARA&A..42...39C,1998AREPS..26..147Z}.
  SiC grains of the X-type are attributed to core-collapse supernovae
  from their large excesses in characteristic isotopes
  \citep{1996ApJ...462L..31N}.
  \Ti-produced overabundance of $^{44}$Ca 
  is found in these SiC X grains, indicating presence of \Ti 
  at their time of condensation. Measured  $^{44}$Ca/$^{40}$Ca ratios are large
    \citep{1996ApJ...462L..31N, 1997Ap&SS.251..355C, 2000M&PS...35.1157H}.
    This proves their supernova origin, on one hand, and likewise it proves
	that dust-producing supernovae may eject \Ti in significant amounts.
	Their attribution to core collapse supernovae rather than 
  supernovae of type Ia \citep{1996ApJ...462L..31N} is plausible, 
  if we believe that
  core collapse supernovae probably dominate \Ti 
  production and 
  because there is
  no direct evidence of dust condensation in thermonuclear supernovae
  \citep{1998ApJ...501..643D}.
  It is likely that all of the observed X grains sample both different 
  condensation environments and different production events. Therefore,
  their $^{44}$Ca abundances cannot be interpreted in 
  absolute terms as a measure of the mass of \Ti ejected per supernova.

   {\it The most plausible cosmic environment for production of \Ti } is
   the $\alpha$-rich freeze-out from high-temperature burning near
   Nuclear Statistical Equilibrium 
   \citep[e.q.][]{1973ApJS...26..231W, 1996ssr..conf...91A}.
   The required high values for the entropy are found in core-collapse
   supernovae.
   Therefore the simplest plausible assumption is that 
   core-collapse supernovae are responsible
   for any substantial sources of $^{44}$Ti. 

   It is likewise plausible that traditional SNIa
   do not add significantly to nucleosynthesis at mass A=44, 
   specifically no $^{44}$Ti,
   since their NSE freeze-out conditions will not be favorable 
   for \Ti production
   \citep{1995ApJS...98..617T}. We consider the symbiotic-star scenario as
   a rare subclass of SNIa, even though their \Ti yields   
   may be large \citep{1994ApJ...423..371W};
   so they indeed would be rare outliers, rather than 
   typical \Ti producing supernovae.
   We will discuss uncertainties in each of these sites in 
   more detail below (see \ref{sec:SNmodel}).

\subsection{Ca from Supernovae} 
\label{sec:ChemEvol}
 
  Measurements of {\it cosmic isotopic abundances} can be converted into 
  isotope production rates; however, for all long-lived isotopes models of 
  chemical evolution have to be applied
  \citep[e.g.][]{1988MNRAS.234....1C, 1997nceg.book.....P, 
  1995ApJS...98..617T, 2003ceg..book.....M}.
  How can we use these for predicting the shortlived \Ti source appearance 
  in gamma-rays?    
  Its lifetime is too short for mean chemical evolution arguments.
  
  Over the time scale of chemical evolution of the Galaxy, the cumulative and
  averaging effects of different explosion types would integrate to a smooth
  pattern of "standard abundances",
  as they are observed throughout the universe.
  Evolutionary pathways for individual elements can be associated
  with the evolution of metallicity and the rates of different supernova types.
  The abundance of $^{44}$Ca in solar-system matter thus
  can be translated through
  models of chemical evolution into a current average production rate of
  \about~3~10$^{-4}$\Msol of \Ti per 100 y
  \citep[][ see below]{1990ApJ...357..638L}.

  Chemical evolution calculations \citep{1995ApJS...98..617T}
  using computed ejecta masses \citep{1995ApJS..101..181W} were
  shown to account reasonably for most solar abundances, including
  specifically $^{40}$Ca, 
  but failing by a factor of three for $^{44}$Ca; this fact
  is also evident from the ratio of \Ti to Fe in cc-SN models.
  But over short time scales where only a limited number of source
  events contribute,
  great variability among sources would
  present a significant difference of the present-day picture from the average.
  Therefore we may compare the expectations from long-term averaged \Ti
  production
  to the present-day \Ti source record imaged in gamma-rays, allowing for the
 short-term  fluctuations with Monte-Carlo realizations of the sources.

 {\it The solar-system $^{44}$Ca abundance} is rooted plausibly in 
 \Ti nucleosynthesis \citep{1973ApJS...26..231W, 1982ena..conf..401C},
 and is a result of the
 integrated Galactic nucleosynthesis prior to formation of the solar system.
 Given a time-dependence of the production rate, it can be
 normalized to the rate required to give precisely the measured solar
 abundance 4.5 Gyr ago, which also fixes the long-term average
 production rate today. Models of galactic chemical evolution,
 constrained by a number of observables, provide us with that
 time dependence, albeit subject to a number of assumptions and
 parameter choices.
 Adjusting the \Ti yield in the aforementioned calculation
 \citep{1995ApJS...98..617T}
 upward by a factor of three to achieve the solar abundance,
 we infer a production rate of
 $^{44}$Ca of about 3.6~10$^{-6}$ \Msol yr$^{-1}$.
 Other considerations using different chemical evolution models
 (see Appendix~\ref{sec:Ca44constraint}) lead from the $^{44}$Ca abundance to a
 current \Ti production rate:
 p($^{44}$Ti)=5.5~10$^{-6}$\Msol~yr$^{-1}$, with full uncertainty
 range 1.2--12~10$^{-6}$\Msol~yr$^{-1}$ (see Appendix
 \ref{sec:Ca44constraint}).

\subsection{Supernova Rates} 
\label{sec:characteristics}

Direct supernova rate measures have been made through correlations between
supernova activity and other tracers of the massive star content of a
galaxy. 
\citet{1991supe.conf..711V}
finds (2.62 +/- 0.8) h$_{100}^{2}$ SN century$^{-1}$. For
h$_{100}$= 0.75, the rate is 1.5 +/- 0.8 SN century$^{-1}$. 
This rate is based on a
combined study of galactic supernova remnants, historical SNe, and novae
in M31 and M33. \citet{1993A&A...273..383C} refer to this rate as the
best estimate. \citet{1991ARA&A..29..363V}
find SNR = 4.0 SN century$^{-1}$. The authors review supernova rates in external
galaxies and derive a specific supernova frequency, in units of 1 SNu =
one SN per century per 10$^{10}$ L$_\odot$(B), for various galaxy types. If one
assumes that the Galaxy is intermediate between types Sab-Sb and types
Sbc-Sd, the specific rate is ~3 h$_{100}^{2}$ SNu. For a Galactic blue-band
luminosity of L(B) = 2.3$\times$10$^{10}$ L$_\odot$(B)  (their Table 11) 
and h$_{100}$ = 0.75 we
infer SNR = 4.0 SN century$^{-1}$. Their review paper also discusses estimates
from internal tracers in the Milky Way: From radio supernova remnant
(RSNR) statistics they infer  SNR = 3.3 +/- 2.0 SN century$^{-1}$. From the
historic record of nearby ($<$ a few kpc) supernovae in the past millennium
they find SNR = 5.8 +/- 2.4 SN century$^{-1}$. 
The large extinction corrections in
the galactic plane make this small sample highly incomplete, which results
in large uncertainties in extrapolations to the full galactic disk. The
authors also review efforts based on the pulsar birth rate, but extensive
observational selection effects in combination with the strong and poorly
understood evolution of luminosity and beaming geometry 
(see \citet{1998puas.book.....L} and \citet{2004hpa..book.....L}) 
renders this method
impractical for estimating the galactic SNR. Continuing the studies of van
den Bergh and Tammann, 
\citet{1993A&A...273..383C}
find
SNR = 1.4 +/- 0.9 SN century$^{-1}$, 
when scaling to external galaxies of similar
type. The sample is obtained from surveys carried out at the Asiago and
Sternberg Observatories. The authors provide an extensive discussion of
the uncertainties of this method, which can exceed 200\% for some late type
galaxies.  More recently, \citet{1994ApJ...425..205V}
find SNR = (2.4-2.7) h$_{75}^2$ events per century. 
This
estimate is based on re-evaluation of the extra-galactic SN rates obtained
from Evans' 1980-1988 observations. This method depends on  a somewhat
uncertain type of the Galaxy and the value of its blue-band luminosity,
while the uncertainty due to the Hubble constant is now very small. Given
the error analysis in the paper, the rate is uncertain by at least 30\%.

\citet{1994ApJ...425..205V} in studying the supernova rates of local
spiral galaxies of types Sab-Sd of
Evans' observations  
estimated that 80\%-90\% of supernova in that galaxies
are of Types Ibc and II. 
Recently \citet{2003sgrb.conf...37C} combining five SN searches
to include 137 SNe in 9,346 galaxies estimates the SN type ratios in the
Galaxy to be Ia : Ib/c : II = 0.22 : 0.11 : 0.67.
From these observations, one infers a ratio of core-collapse to 
thermonuclear supernova of R=(II+Ibc)/Ia = 3.5.
However, note that the Galactic historical record in the last millennium 
shown in Table~\ref{histrec} contains only two Type Ia SNR out of
six SNRs.
An often used alternative distribution over types is
Ia : Ib/c : II = 0.1 : 0.15 : 0.75 
\citep{1997MNRAS.290..360H, 1994JRASC..88..369D, 1994ApJS...92..487T,
1993A&AS...97..219H}, 
which implies a three times higher cc-SN fraction.
In this work, we adopt this set of parameters, and note that
the small $^{44}$Ti yield of Type Ia renders our results insensitive
to this ratio.

Over the last millennium, the historic record contains six events 
(see Appendix \ref{sec:SNrates}), which
  implies a rather low rate at face value. However Galactic extinction
  at visible wavelengths
  and embedded supernovae will lead to large occultation bias, and with
  extinction models plus Monte Carlo simulations this historic record
  can be assessed to approximately agree with extragalactic rate determinations
  (see Section \ref{sec:resolvingconflict} and Appendix \ref{sec:models}).
  An often cited rate of galactic cc-SN of three per century is consistent both 
  with astronomical arguments \citep{1990ApJ...359..277V}
  and with the rate 
  inferred by \citet{1995ApJS...98..617T} 
  from their chemical evolution model that 
  produces solar abundances successfully. 
  We adopt a supernova recurrence rate of 30 years as a baseline for our
  \Ti sky expectations.  
  
\onecolumn
\begin{figure*}[t]
\centering
 \includegraphics*[width=0.9\textwidth]{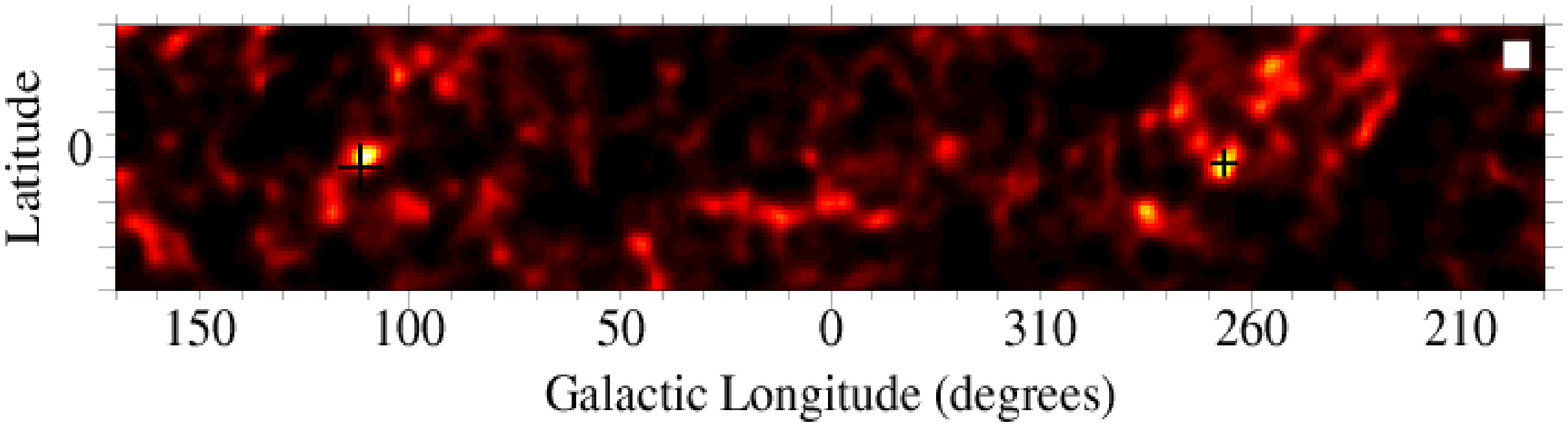}
   \caption{\it Maximum-entropy map of the Galactic plane 
    (within latitude $\pm$30$\degre$)  in the
  $^{44}$Ti energy window (1.066-1.246 MeV) for the combination
  of CGRO observations from 0.1 to 617.1. 
  Two crosses mark the positions of Cas A and GROJ0852-4642.
  This figure is adopted from \cite{1999asra.conf...65I}.}
  \label{fig:timap_IyudinRingberg}
 \includegraphics*[angle=90,width=1.0\textwidth]{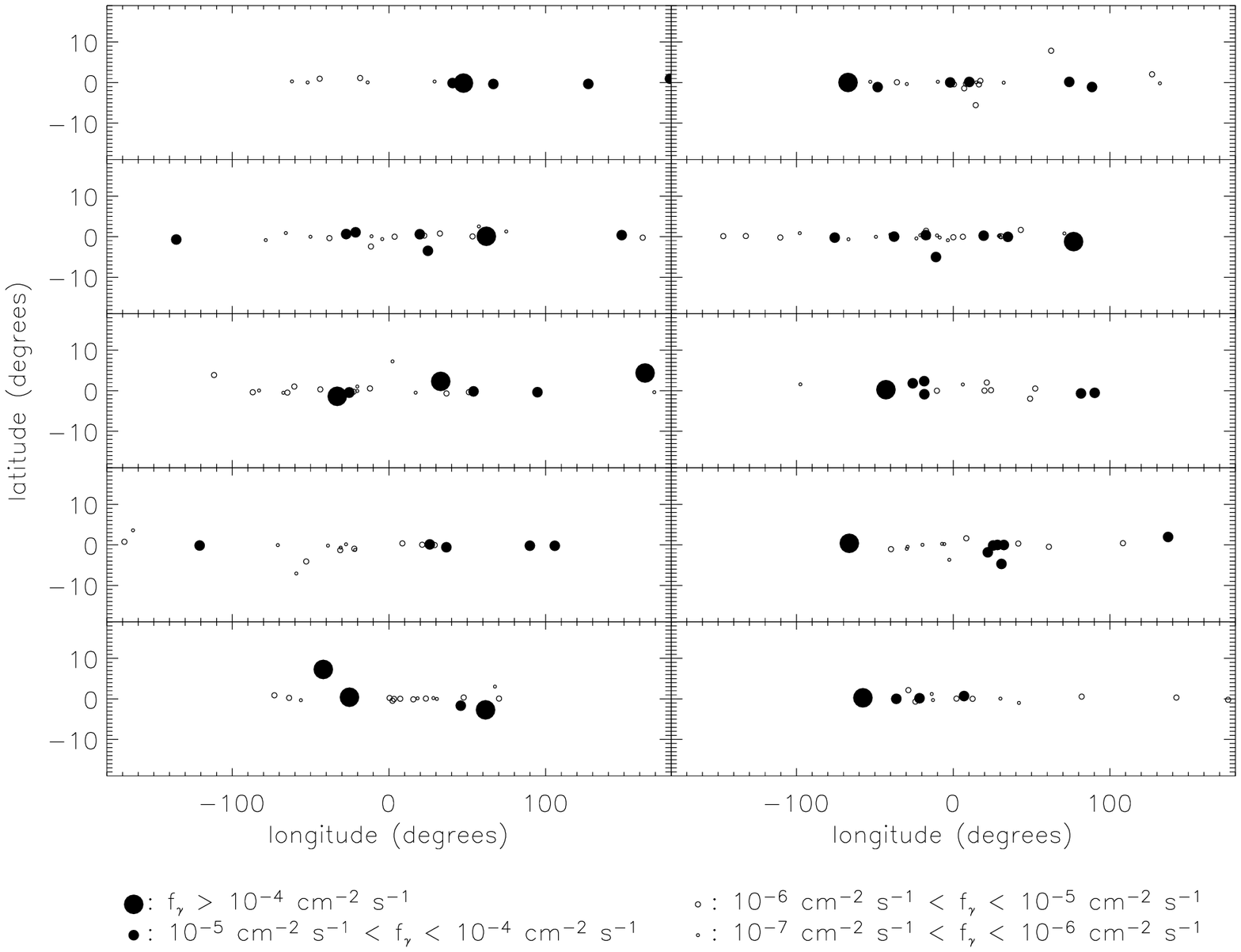}
   \caption{\it The expected Ti sky of 10 simulated galaxies of model {\bf A}
   where the supernova recurrence time is taken to be 30 years and 
   the supernovae ratio of Ia:Ib:II = 0.10:0.15:0.75. 
  Simulating a 10$^5$ galaxy sky,
  a gamma-ray detector with a detection limit of $10^{-5}cm^{-2}s^{-1}$
  would have a probability of detecting 0, 1, and 2  \Ti sources
  of 0.0017, 0.012, and 0.037, respectively. 
  A slightly better instrument than the $10^{-5}cm^{-2}s^{-1}$ detection limit
  would detect several \Ti sources.
  }
  \label{fig:Expected_ti44map}
\end{figure*}

 \twocolumn

\subsection{Supernova Locations}
\label{sec:supernovalocations}
   
     The 1.157 MeV $\gamma$-ray line following 
   \Ti decay has been detected in the 340-year old Galactic supernova remnant 
   Cas A \citep{1994A&A...284L...1I, 2001ApJ...560L..79V}.
   COMPTEL's survey \citep{1997A&A...324..683D, 1999ApL&C..38..383I}
   has resulted in other candidate sources, such as the so-called 
   Vela junior SNR \citep{1998Natur.396..142I}.
   INTEGRAL's inner-Galaxy survey has been studied with IBIS Imager data,
   which did not reveal a new source in this region 
   \citep{2004sf2a.confE..90R}.
  The difficulties of MeV observations
   have thus not led to convincing new \Ti rich 
   supernova remnants, especially in the inner Galaxy region 
   (l=0\degree$\pm$30\degree) 
   where observations are deepest.
   Yet, Cas A seems an established \Ti detection in COMPTEL 
   \citep{2000AIPC..510...54S}
    and Beppo-Sax \citep{2001ApJ...560L..79V} 
	and INTEGRAL/IBIS \citep{2005AdSpR..35..976V} measurements, 
   while OSSE \citep{1996A&AS..120C.357T} 
   and RXTE \citep{2003ApJ...582..257R} measurements of Cas A 
   were not sufficiently sensitive.
   There exist no well-understood supernova remnants other than Cas A 
   where the \Ti production issue can be tested. 
   It is apparently the only supernova whose 
   yield, age, and nearness makes
   \Ti visible in gammas.
   Naturally we ask ourselves if Cas A is 
   a typical supernova or an anomalous case of a high-\Ti yield
   supernova? 

   Current supernova models predict an amount of \Ti 
   which is of the same 
   order than what these observations suggest, though generally slightly less.
   Can we take this as a satisfactory confirmation of our understanding of core 
   collapse supernova \Ti production?
   These two identified core collapse events and their association 
   with \Ti production appear to be
   in line with models which attribute \Ti production to the more 
   frequent standard core collapse
   events but not to standard thermonuclear supernovae. Is this correct? 
   
   If true, the location of \Ti sources should match the locations 
   of young massive stars which have rather short lifetimes. 
   There is substantial evidence that massive star formation occurs in spiral
   arms and predominantly in the inner Galaxy
   \citep{2003ApJ...590..271E, 2001AJ....122.3017S, 1998ARA&A..36..189K}.
   Massive stars can be observed directly in the infrared 
   \citep[e.g.][]{1994ARA&A..32..227M},
   though extinction corrections are large in regions of dense clouds. 
   Possibly an even better (though more indirect) massive-star census 
   can be derived
   from $^{26}$Al decay $\gamma$-rays 
   \citep{1995A&A...298..445D, 1996PhR...267....1P, 2000NewAR..44..315K,
   Diehl2005}. 
   $^{26}$Al  is understood to originate 
   predominantly from massive stars and $\gamma$-rays 
   easily penetrate even dark clouds in star forming regions.
   So, do we see the \Ti sources in regions where we expect them to occur?
   Or do other factors which are not yet understood conspire to make \Ti
   ejection a phenomenon of core collapse events occurring in special
   regions and environments?

\subsection{Estimating the \Ti Sky Appearance}
\label{sec:Tiexpectations}

If we want to estimate how the \Ti sky should appear in a gamma-ray survey,
we need to follow a statistical approach, due to the rare occurrence
of supernovae. We therefore apply a Monte Carlo approach of sampling
plausible probability distributions for supernova rates, their \Ti yields,
and their Galactic distribution, thus calculating a large statistical sample
of possible appearances of the \Ti sky. 

\begin{figure*}[t]
 \includegraphics[angle=90,width=0.95\textwidth]{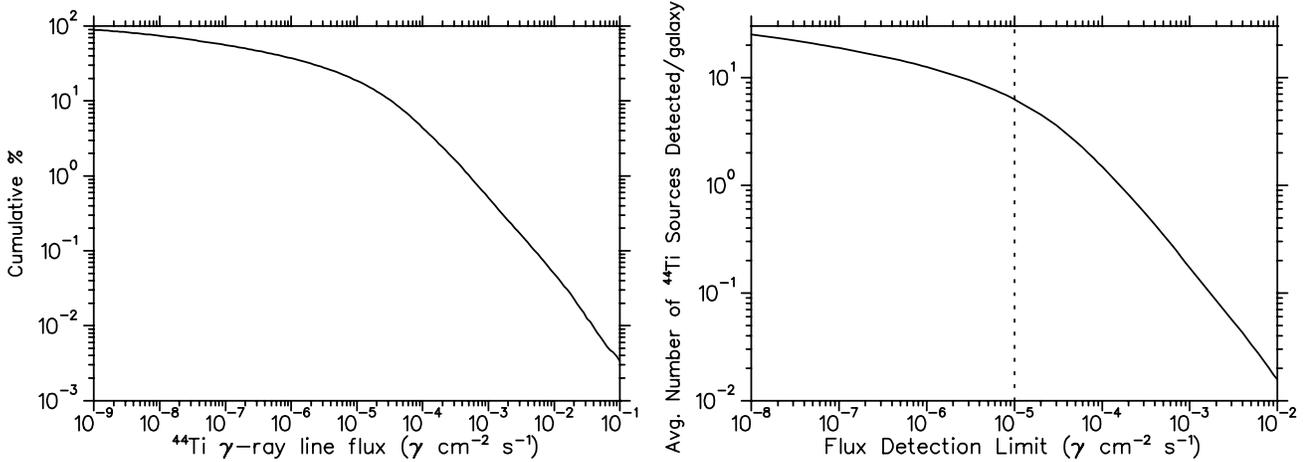}
   \caption{\it Cumulative \Ti gamma-ray line flux distribution of
            supernovae with $f_{\gamma}>$1$\times$10$^{-12}$ cm$^{-2}$s$^{-1}$ 
            according to our
            Model {\bf A} (see Appendix \ref{sec:models})
            with supernovae type ratio 
            of Ia:Ib:II = 0.1:0.15:0.75 (left).
            Expected average number of supernovae per sample galaxy
            with their $^{44}$Ti gamma-line fluxes above detection limits,
            for a supernova recurrence time of 30 yrs, irrespective of their 
	    position in the sky (right). The detection limit of COMPTEL
            instrument is shown as a dotted line.}
  \label{fig:CumulativeGflux}
\end{figure*}

\begin{itemize}
\item 
The mean rate of supernovae in our Galaxy is taken as one event in 30 years.
\item
\Ti\-producing supernova events can be of Type Ia, Ib/c, or II SNe.
    For this study, we choose a supernova type 
    ratio, Ia~:~Ib/c~:~II = 0.10~:~0.15~:~0.75 
\citep{1994JRASC..88..369D, 1997MNRAS.290..360H}.
\item
Supernovae are then assumed to produce \Ti according to models of that type.
In our models,
Type II supernova yields are generated according to stellar mass, which
is drawn from a Salpeter-type initial mass distribution 
for M $\ge$ 8 M$_{\sun}$;
the $^{44}$Ti yields per mass of the star are taken from 
Table 1 of \citet{1996ApJ...464..332T}. 
A typical yield is 3$\times$10$^{-5}$ M$_{\sun}$ for M=25~M$_{\sun}$,
but variations with mass are about a factor~2.
The $^{44}$Ti yields of Type Ib SNe are uniformly distributed between
3$\times$10$^{-5}$ M$_{\sun}$ and 9$\times$10$^{-5}$ M$_{\sun}$
with typical values of 6$\times$10$^{-5}$ M$_{\sun}$ 
\citep{1996ApJ...464..332T}.
The $^{44}$Ti yields of Type Ia SNe are uniformly distributed between
8.7$\times$10$^{-6}$ M$_{\sun}$ and 2.7$\times$10$^{-5}$ M$_{\sun}$
so this range covers the $^{44}$Ti yield in the deflagration W7 model,
the delayed detonation WDD2 model, and the late detonation W7DT model
of \citet{1997NuPhA.621..467N}.
But instead of adopting the \Ti ejecta masses directly from these models, 
we uniformly increase these by a factor 3 in order 
that chemical evolution calculations reproduce the known solar
$^{44}$Ca/$^{40}$Ca ratio. This
assumes that  \Ti and $^{44}$Ca nucleosynthesis are directly related and
that some unknown physics factor causes the computed \Ti masses
to be uniformly low by a factor three.
Note that \Ti yields and supernova rate determine the brightness scale of
our \Ti sky expectations.
\item
For each supernova type, we then adopt a parent spatial distribution, from
which we draw random samples to determine the location of the event.
The spatial distribution for core-collapse supernovae (Types Ib/c and II) is
taken, from either an exponential disk with scale radii between 3.5 and 5 kpc, 
or from a Gaussian-shaped ring at 3.7~kpc radius with a thickness of 1.3~kpc,
both placing young stars in the inner region of the Galaxy as suggested by many
observables.
Thermonuclear supernovae (Type Ia) are assumed to arise from an old stellar
population, hence we adopt the distribution of novae as our parent Galactic
distribution for those rare events. We adopt the composite disk-spheroid
nova model from \citet{1987ApJ...317..710H} 
(See Appendix \ref{sec:montecarlo} 
for more detail on the variety of modeled distributions).
\item 
We then randomly choose the individual supernova age within the last
millennium, for deriving its \Ti decay gamma-ray brightness. 
 This assumes that the factors governing the mean recurrence rate have
not changed in the past 10$^{3}$ yr.
\end{itemize}

Randomly selected sample results from 10$^{5}$ 
such realizations of an expected \Ti sky are shown in
Fig. \ref{fig:Expected_ti44map}, to be compared with 
the observed 1.157 MeV COMPTEL image shown in 
Fig. \ref{fig:timap_IyudinRingberg} \citep{1999ApL&C..38..383I}.
We illustrate how typical these expected images would be, by
showing the distribution of source brightnesses over a much larger number of
Monte Carlo samples (Fig. \ref{fig:CumulativeGflux}, left): A \Ti flux
above a representative limit of 10$^{-5}$ph~cm$^{-2}$s$^{-1}$ occurs in 
19\% of our \Ti gamma-line flux distribution.
The expected number of \Ti point sources per galaxy
lies above the gamma-ray survey sensitivity limit,
 i.e., we do expect typically 5-6 positive detections of \Ti sources
(see Fig. \ref{fig:CumulativeGflux} right).

\begin{figure}[htp]
 \includegraphics[angle=90,width=0.5\textwidth]{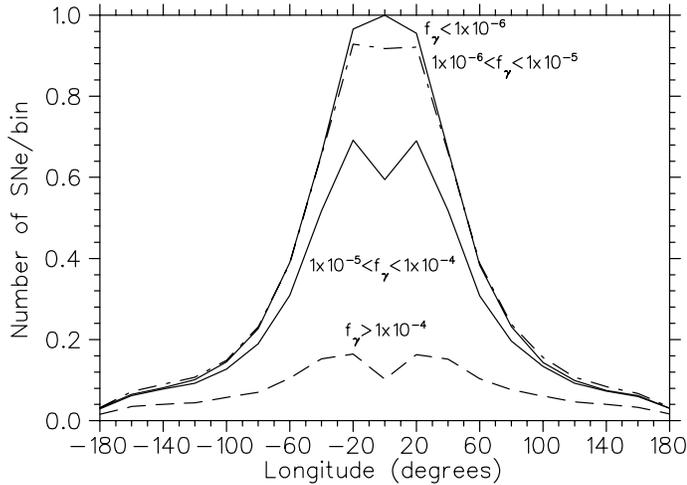}
\caption{\it The longitude distribution of supernovae 
for four $^{44}$Ti gamma-ray line flux bandwidths 
in Model {\bf A} of 
Appendix \ref{sec:models} 
with supernovae type ratio of Ia:Ib:II = 0.1:0.15:0.75.
The equal-bin width is 20 degrees in longitude. 
Only the $f_{\gamma}<$1$\times$10$^{-6}$ cm$^{-2}$s$^{-1}$ distribution 
is normalized and multiplied by a factor of 0.298.}
\label{fig:longdist}
\end{figure}

It is immediately evident from Figures \ref{fig:Expected_ti44map}
and \ref{fig:CumulativeGflux} that expectations based on
such seemingly plausible assumptions look very different than 
the observed \Ti sky: 
Fig. \ref{fig:Expected_ti44map} 
shows $\sim$~4--7 observable \Ti sources in an area of the Galaxy that contained
none (see Fig. \ref{fig:timap_IyudinRingberg})
 above the observable flux limit used for comparison. The brightest
source in that realization (large-filled circles)
has I$_\gamma$~$\sim$~10$^{-4}$cm$^{-2}$s$^{-1}$, 
which would have been seen as a {\it $\sim$10~$\sigma$} source by COMPTEL,
and would have been detected already in 
INTEGRAL's inner-Galaxy survey \citep{2004cosp.meet.4023V}. 
The majority of our calculated samples lead to this same type of conflict
(see Fig. \ref{fig:CumulativeGflux}).
The probability of having no sources within the
central galaxy is small for current surveys, as illustrated in 
Fig. \ref{fig:longdist} (for model A):
For a survey down to I$_\gamma$~$\sim$~10$^{-5}$cm$^{-2}$s$^{-1}$, 
$\sim$12\% of supernovae appear within  longitudes 0$\pm$60$^\circ$,
whereas only 72\% of supernovae is within that volume .
An interesting feature, however, is that the longitude
distribution for $f_{\gamma}<$1$\times$10$^{-5}$ cm$^{-2}$s$^{-1}$ 
is dominated by sources from Type Ia (the bulge) and farther away SNe,
while the distribution for 
$f_{\gamma}>$1$\times$10$^{-5}$ cm$^{-2}$s$^{-1}$ 
is dominated by sources from Type Ib and Type II (disk) and nearby SNe.
This feature also can be seen in Fig. \ref{fig:CumulativeGflux}, 
where around 
$f_{\gamma}\simeq$1$\times$10$^{-5}$ cm$^{-2}$s$^{-1}$
the distribution changes its slope.

It is clear that something is wrong with at least
one of these assumptions: 
   \begin{itemize}
   \item the rate of core collapse supernovae in the Galaxy is 3/100~y
   \item a core collapse supernova produces \about 10$^{-4}$\Msol of 
         \Ti 
   \item supernovae of each given type sample a relatively narrow range
         of the \Ti yield, i.e., we do not consider extreme outliers event 
         (nor do we consider a rare but superhigh yield source class). 
   \item the locations of core collapse supernovae is axisymmetric
         i.e., we assume star formation in spiral arms is not dominant.
   \end{itemize}

In the following we examine these questions, and seek 
possible explanations for the apparent conflict.

\section{Resolving the \Ti Sky Conflict}
\label{sec:resolvingconflict}

   \subsection{Gamma-ray Observations}
   \label{sec:gammaraydata}
		 
The \Ti gamma-ray sky can be studied in lines from
the primary decay to $^{44}$Sc at 67.9 and 78.4~keV and
from the following decay to $^{44}$Ca at 1.157~MeV.
In this latter line, the COMPTEL imaging telescope had
reported the pioneering detection of \Ti from Cas~A
\citep{1994A&A...284L...1I},
clearly showing a point source image at 1.157~MeV as well as the line in a
spectrum from this source.
This detection
 had initially created some controversy because other gamma-ray 
 instruments apparently
 did not see it \citep{1996A&AS..120C.357T, 1998axrs.symp...68R}.
 We now believe that this is due to the high initially-reported 
 COMPTEL gamma-ray flux value, reduced later
 with better statistical accuracy of the measurement
 \citep{1999ApL&C..38..383I, 1997A&A...324..683D}. The independent
 detections with BeppoSax \citep{2001ApJ...560L..79V} and with
 INTEGRAL/IBIS \citep{2005AdSpR..35..976V}
 in both lower-energy lines (Beppo-SAX) and in the 67.9~keV line (IBIS)
 now consolidate the \Ti detection from Cas A,
 but also suggest that indeed  the \Ti gamma-ray flux of Cas A
 is in a range between
 0.8 and 3.5~10$^{-5}$~ph~cm$^{-2}$s$^{-1}$; an ``average'' of
 (2.6$\pm$0.4$\pm$0.5)~10$^{-5}$~ph~cm$^{-2}$s$^{-1}$ has been derived
 \citep{2005AdSpR..35..976V}.
 An upper limit from INTEGRAL/SPI reported from first studies
 is consistent with this flux value, and may suggest that the \Ti line
 is broader than 1000~km~s$^{-1}$
 \citep{2005AdSpR..35..976V}.

 COMPTEL's sky survey allowed for mapping of the plane of the Galaxy in
the \Ti line (Dupraz et al. 1996, Iyudin et al. 1999). 
 Secondary features in these COMPTEL \Ti maps kept the discussion about
 statistical significances and systematic uncertainties alive 
 (see \citet{2000AIPC..510...54S}  for a comparison of Cas A to RXJ0852, 
 a promising second source candidate,  \citet{1998Natur.396..142I}).
 The COMPTEL point source detection algorithm
 \citep{1992daia.conf..241D} has been
 tested with simulations
 over the full sky: Likelihood statistics has been verified to reproduce the
 expected number of artificial sources for a full sky survey, as the noise
 level is approached.
 The problem is that in the range of all the \Ti gamma-ray lines,
 all gamma-ray telescopes suffer from a large background
 from local radioactivity induced by cosmic-ray
 bombardment of the instruments
 \citep{1999APh....11..277G, 2002NewAR..46..625W}.
 Determination of this background is crucial.
 For an imaging instrument, this can be done rather well by interpolation
 of imaging signatures from adjacent energies.
 But furthermore, for COMPTEL the 1.157~MeV
 line of \Ti is not far above its lower energy threshold, and in fact
 imaging selections
 strongly affect the sensitivity of the instrument up to \about~1.5~MeV.
 Nevertheless, imaging analysis in adjacent energy bands should all experience
 similar problems, and therefore differences between images in the \Ti band and
 in neighboring energy bands can be attributed to \Ti rather than
 continuum sources
 or instrumental artifacts, once they are confirmed to be point-like sources
 (instrumental background lines would in general spread over data space and
 hence lead to extended or large-scale artifacts in the image, hence impact
 on the flux measurement rather than on point source detection).  Detailed
 comparisons of results for different energy bands, data subsets,
 and selections
 have led to the more cautious report about the Vela-region source
 \citep{2000AIPC..510...54S},
 so that still no convincing second \Ti source clearly above
 the noise level is claimed.
  The INTEGRAL limits for \Ti lines from the Vela region source 
  are now close to the 
 reported COMPTEL flux value for this candidate source
 \citep{2004ESA-SP.552R, 2004ESA-SP.552K}
 and we therefore consider only one single \Ti source (Cas~A) as being
 detected down to
 flux levels of  ~10$^{-5}$~ph~cm$^{-2}$s$^{-1}$.

 For the low-energy lines measured by other instruments, 
 in addition to the instrumental
 background the underlying continuum 
 (see \citet{1999asra.conf...77T} for Cas A)
 presents a major source of uncertainty 
(see \citet{2001ApJ...560L..79V, 2003ApJ...584..758V, 2005AdSpR..35..976V}).
 This becomes even more of a problem for recombination line 
 features from $^{44}$Sc
 at 4.1~keV, which have been studied with ASCA 
 \citep{2000PASJ...52..887T, 2005A&A...429..225I},
 yet without clear detections 
(though tantalizing hints have been discussed for the Vela-region source).

 For our quantitative comparisons of the \Ti supernova rate with the historical
 record of supernovae in the last millennium 
 (see Appendix \ref{sec:fluxdataspace}), 
 we make use of the COMPTEL survey.
 We avoid the regions where reduced exposure  might eventually
 lead to increased artifact levels, and concentrate on the inner Galaxy where
 exposure for the COMPTEL sky survey is deep and homogeneous. We also avoid 
 using data after the second re-boost of the satellite, whereafter 
 the activation
 level of $^{22}$Na had increased substantially 
\citep{2002NewAR..46..625W}.
 This leaves us with a dataset covering the sky range of 
 $|l|\leq$90$\degre$, $|b|\leq$30$\degre$ using
 7 years of the 9-year sky survey. In these gamma-ray data, the \Ti source
 sensitivity should be rather well-behaved and useful for our study.
 Note that the first inner-Galaxy survey from INTEGRAL 
 (longitudes of~$\pm$20$^o$ around the GC)
 \citep{2004ESA-SP.552R}
 is consistent with our dataset in that also there no source is found
 to flux levels of  ~10$^{-5}$~ph~cm$^{-2}$s$^{-1}$
  \citep{Renaud2005}.

\subsection{Galactic Supernova Rate}
\label{sec:GalacticSN}

Many determinations of the galactic star formation rate (SFR) and
supernova rate (SNR) have been made in past decades (see Diehl et al. 2005
for a compilation of estimates, and \citet{2005fost.book.....S} on general
astrophysical aspects of star formation).  In this work we are concerned
with the $^{44}$Ti from supernovae, so we need the formation rate of
massive stars (above, say, 10 M$_\odot$). The conversion between SNR and
SFR is a sensitive function of the Initial Mass Function (IMF),
and values given in the literature thus vary depending on the author's
chosen IMF. For a generic transformation equation, we use the
calibration from \citet{1997ApJ...476..144M}: SFR = 196 SNR, where the SFR
is measured in M$_\odot$ yr$^{-1}$ and the SNR in events per year. A
star formation rate of 4 M$_\odot$ yr$^{-1}$ thus corresponds to a
supernova rate of two events per century. These rates only include core
collapse supernovae (Type II and Ibc), but not SNIa.
Generally these supernova rates are averages over 
time scales much longer than \Ti decay. 

Many papers discuss the star formation (rate) history (SFH) in relative
terms, or the star formation rate surface density 
(M$_\odot$ yr$^{-1}$ kpc$^{-2}$) 
in the solar neighborhood and its radial dependence. None of
these papers is useful (for our purpose) without  an absolute calibration,
based on a model of the galactic distribution of star formation.

Generically, the SFR is obtained from a tracer that can be corrected for
observational selection effects and is understood well enough so that
possible evolutionary effects can be taken into account. One either deals
with a class of residual objects, such as pulsars or supernova remnants,
or with reprocessed light, such as free-free, H-alpha, or IR emission that
follows from the ionization and heating of interstellar gas and its dust
content in the vicinity of the hot and luminous stars. One must be careful
to include time-dependent effects, as the afterglow of an instantaneous
starburst behaves differently than the steady state output from a region
with continuous star formation. Here, we are concerned with an average
star formation rate.

To set the stage, let us collect SFR values from the literature.
\citet{1978A&A....66...65S}
concluded
SFR = 5.3 M$_\odot$ yr$^{-1}$, 
while \citet{1980ApJ...235..821T} finds a very low
value SFR = 0.8 M$_\odot$ yr$^{-1}$, and \citet{1982VA.....26..159G} 
report a very high value of SFR = 13.0 M$_\odot$ yr$^{-1}$. Later papers
appear to converge on a median value: \citet{1984VA.....27..303T}
finds SFR = 3.0 M$_\odot$ yr$^{-1}$, and \citet{1987sbge.proc....3M}
finds 5.1 M$_\odot$ yr$^{-1}$. Measurements in the past
decade confirm this moderate rate:  \citet{1989ApJ...345..782M}
derives SFR = 3.6 M$_\odot$ yr$^{-1}$ 
from the analysis of thermal radio emission
(free-free) from HII regions around massive stars. This emission is
directly proportional to the production rate of ionizing photons, which in
turn is directly proportional to the SFR. He pointed out that this method
is very sensitive to the slope of the high-mass IMF. We also note that the
method depends on stellar model atmospheres in conjunction with models for
massive stars, which change with treatments of mass loss, rotation, and
convection. This paper also briefly discusses the use of the far-IR
luminosity, due to warm dust heated by the absorption of photons from
massive stars. The author uses the measured IR luminosity of the Galaxy of
4.7$\times$10$^9$ L$_\odot$ (from \citealt{1987sbge.proc....3M})
to derive SFR = 2.4 M$_\odot$ yr$^{-1}$.

We must convert between star formation rate (SFR) and supernova rate (SNR) (see above). 
\citet{1997ApJ...476..144M}
promote the value SFR = 4.0 M$_\odot$ yr$^{-1}$,
and based on the 
Scalo's IMF \citep{1986IAUS..116..451S} convert this rate into a total number
rate of 7.9 stars per year. They assume that all stars above 8 M$_\odot$ become
supernovae, corresponding to a supernova fraction of 2.6$\times$ 10$^{-3}$. The
mean stellar mass is $<m>$ = 0.51 M$_\odot$. The corresponding cc-SN rate is 2 per
century. This value is also supported by a completely independent method,
based on the production of radioactive $^{26}$Al in cc-SNe, which can be
traced though its gamma-ray line at 1.809 MeV. 
\citet{1997ApJ...479..760T}
use this method and conclude
SFR = (5 +/- 4) M$_\odot$ yr$^{-1}$, utilizing the $^{26}$Al line flux derived from COMPTEL.
The steady-state mass of $^{26}$Al obtained in their work is in the
range 0.7 - 2.8 M$_\odot$, consistent with the value presented in the
recent study by Diehl et al. (2005). Based on the Salpeter IMF in the
range 0.1 - 40 M$_\odot$ and the $^{26}$Al yields 
from \citet{1995ApJS..101..181W} 
[which do not include contributions from the Wolf-Rayet wind
phase] the authors derive the above quoted SFR and an associated
cc-supernova rate of 3.4 +/- 2.8 per century. They neglect 
hydrostatically produced $^{26}$Al that is injected into the ISM in massive
star winds, which causes their SFR to be overestimated. The large uncertainty is
mostly due to the steady-state mass of $^{26}$Al inferred from the COMPTEL flux.
INTEGRAL data presented by Diehl et al. (2005) have significantly reduced
the error in this key quantity, and with wind yields included the latter
study finds SNR = 1.9 $\pm$ 1.1 supernovae per century, which corresponds
to SFR = 3.8 $\pm$ 2.2 M$_\odot$ yr$^{-1}$, similar to the value
given in \citet{1997ApJ...476..144M}. This SFR is very similar to the one
obtained for M51 \citep{2005ApJ...633..871C}, and thus places the
Galaxy in the group of quiescently star-forming galaxies.

The most recent paper addressing this issue is by \citet{2005AJ....130.1652R},
who does not derive the SFR, but states that
the galactic supernova rate is estimated as probably not less than 1
nor more than 2 per century. Using the conversion factors from 
\citet{1997ApJ...476..144M},
one infers that the SFR is in the range 2-4 M$_\odot$ yr$^{-1}$. Reed
uses a sample of a little over 400 O3-B2 dwarfs within 1.5 kpc of the Sun
to determine the birthrate of stars more massive than 10 M$_\odot$. The galaxy
wide rate is derived from this local measurement by extrapolation based on
models for the spatial distribution of stars, a model for galactic
extinction (to accomplish corrections for stellar magnitudes), and a model
of stellar life times. Reed emphasizes various sources of errors, such as
lacking spectral classifications of some bright OB stars, the (poorly
known) inhomogeneous spatial structure of extinction as well as stellar
density, and non-unique connection between mass and spectral type.
Finally, Reed also draws attention to the fact that one would have to
include B3 dwarfs as well, 
if the lower mass limit for supernovae is 8 M$_\odot$ 
and not 10 M$_\odot$ 
(see \citet{2003ApJ...591..288H} for comments on this mass limit). The
OB-star catalog of the author was used to perform a modified V/Vmax test
to obtain a present-day star count as a function of absolute V-band
magnitude. From the stellar life times, and the assumption of steady
state, the local birthrate follows. A double exponential model (in
galactocentric radius and scale height above the plane) of the spatial
distribution of these stars (which includes an inner hole of radius R =
4.25 kpc) ultimately leads to a total birthrate of 1.14 OB stars per
century. Variations in the size of the hole change this number
significantly, which leads the author to finally claim a rate of 1-2
supernovae per century.


In the context of our interpretation of the $^{44}$Ti observations, we
would argue that
Supernova rates between one and three cc-SNe per century are
consistent with the large set of studies reviewed above. To solve the
$^{44}$Ti Sky conflict with a choice of the SNR (or SFR), an extremely low
rate outside this range would of course explain the absence of $^{44}$Ti gamma-ray
line sources in the sky. However, chemical evolution arguments for
$^{44}$Ca would then require correspondingly higher $^{44}$Ti yields which
are not supported by explosive nucleosynthesis studies (as discussed
below) and which in any case would lead to higher fluxes from supernova
remnants and thus again to a source count that exceeds the observed count
of $\sim$ 1. The observational constraints are on the product of rate and
yield. The natural solution to the problem may
be the very rare events with extremely high yields, which is discussed below.
In that case it is of course totally unclear what to use for the spatial
distribution of these events, and ``unusual'' positions of a gamma-ray line
source on the sky (such as that of Cas A) would be hard to interpret.

But perhaps we have overlooked another option. We know that the galactic
star formation process is strongly correlated in space and time 
\citep{2003ApJ...590..271E}.
Could it be that the Galaxy just had a brief hiatus in
its SFR? It would not take too much of a pause (say a few centuries) to
explain the absence of bright $^{44}$Ti sources if the past few centuries
were very untypical with respect to the SFR (or SNR). This possibility is included in
our Monte Carlo simulations, which show that this is not a likely solution
when one simply considers Poisson fluctuations. This solution is thus not
acceptable, unless one can point out a physical cause of the hiatus in the
recent SFR.

\subsection{Galactic Supernovae Locations}
\label{sec:GalacticSNLoc}

Likewise, we may wonder about the possibility of very large
spatial fluctuations. Could it be that the recent Galaxy exhibits an
average star formation, rate-wise, but a lopsided distribution in space.
If the opposite side of the Galaxy currently forms stars, and regions in
the solar sector are relatively inactive ($l=90-270\degree$), 
we expect to detect fewer
gamma-ray line sources (because of their somewhat larger average
distances). But at the same time the supernovae would also suffer from
enhanced extinction, and matching the historic SN record would require a
much higher rate. We have simulated the effects of a lopsided  profile
with a von Mises Distribution, the analog of a Gaussian distribution for
circular data \citep{1975JRSS.37..349M,1996sacd.book.....F}. 
This function allows us to change from an
axis-symmetric galactic distribution to a one-parameter distribution
(measured by a parameter k, where k=0 corresponds to the
uniform circular distribution) in a chosen direction 
which we choose it to be the longitude $l=0\degree$ direction. 
We re-simulated the
10$^5$ Galaxies sample as described in Sec. \ref{sec:Tiexpectations}, 
and find that the
detection probabilities 
(stated in the caption of Fig. \ref{fig:Expected_ti44map}) change to 
0.0042, 0.023, and 0.062 for parameter k=0.5 and
0.0068, 0.035, and 0.087 for parameter k=1.0, respectively.
The probability curves for $^{44}$Ti $\gamma$-line source detection
(see Fig. \ref{fig:ExpDisk}) are shifted towards higher rates, as expected, but the
overall likelihood of these models decreases. The constraint from the
historic record demands new rates that are even larger, as the
extinction correction affects the results more strongly than the D$^{-2}$ distance
effect for the flux. The combination of these two constraints make
lopsided models less acceptable than axis-symmetric ones.

A lopsided model would also make Cas A even more special, regarding its
unexpected location on the sky. To alleviate this problem we simulated
lopsided star-forming galaxies in which the solar sector was the more
active.  
The detection probabilities of 0, 1, and 2 $^{44}$Ti sources of
this model are
0.0005, 0.003, and 0.012 for parameter k=1.0, respectively.
This shows that the number of 
the most probable $^{44}$Ti $\gamma$-line source detection
in the Galaxy is $>$2 detections
and the model is less probable than the model used for 
Fig. \ref{fig:Expected_ti44map} for consistency with the observed \Ti sky.

Is it reasonable at all to consider one-sided star forming galaxies? That
major merger events should be able to tidally induce a lopsided
starburst activity is perhaps obvious, but the Galaxy is not undergoing
such an event. However, \citet{2000ApJ...538..569R}
have shown that even minor mergers \citep{1995MNRAS.277..781I} 
generically termed ``weak interactions'' may lead to a boost in the star
formation rate correlated with their lopsidedness. Another mechanism for
the creation of non-symmetric star formation patterns is the interaction
between odd numbers of spiral density waves, as it may be at work in M51
\citep{2003AJ....126.2831H}.
We do not advocate such an
asymmetry for our Galaxy, but just wanted to consider this real
possibility as one of the potential fixes for the $^{44}$Ti sky problem.
Our simulations indicate that even such an extreme solution does not work,
as the various combined constraints operate against each other. While a
lopsided Galaxy helps on the gamma-ray source count side, the historic
record is harder to explain if recent supernovae are located
preferentially on the far-side.

\subsection{Galactic Extinction Map}
\label{sec:GalacticExtMap}

   Analyzing the consistency of the supernova rates derived from the
Galactic historical record and the COMPTEL's gamma-ray map 
(Appendix~\ref{sec:SNrates}), 
the consistency would improve if the SN rate from historical record
is smaller. This could be realized if the true Galactic extinction map is 
lower than the visual extinction map we used here.
Recently there are two published Galactic extinction maps that are
useful for the type of study of this paper.
The optical reddening model of \citet{1998A&A...330..910M},
which is based on Galactic dust distribution model,
makes use of the same 
optical sky surveys implemented by \citet{1997AJ....114.2043H}
in addition to some other restricted surveys. However this extinction map is 
reliable only for solar neighborhood within 6 kpc.
Another large scale three-dimensional model of Galactic extinction
based on the Galactic dust distribution of \citet{2003A&A...409..205D}
has been shown to give a good agreement with the empirical extinction
derived from NIR color-magnitude diagrams within 0.05 magnitude and 
furthermore it is reliable for a distance up to $\simeq$8 kpc. 
This extinction model gives a larger magnitude of
extinction than that of \citet{1997AJ....114.2043H} for 
longitude $|l| \leq$ 1.5\degree and 
for most pointing directions from the Sun
for distance larger than 6 kpc. For distance less than 5 kpc,
the extinction of this map is smaller than that of
\citet{1997AJ....114.2043H} which could give a better agreement 
between the SN rates of the historical record and 
of the  COMPTEL's gamma-ray map.

\subsection{The SN model}
\label{sec:SNmodel}
		
\subsubsection{Lifetime of $^{44}$Ti}
\label{sec:ti44lifetime}

Although the lifetime of $^{44}$Ti measured in laboratories
had exhibited a 
large uncertainty since its first measurement in 1965, a compilation of
five recent experiments 
performed after 1998 gives an averaged lifetime
of  87 $\pm$ 1 yr,      
where the quoted error is of statistical
and of one standard deviation 
(see, e.g., Figure~5 of \citet{2001NuPhA.686..591H}
and also \citet{1998PhRvL..80.2554G}).
Apparently, this small uncertainty in the measured $^{44}$Ti lifetime does
not affect the discrepancy discussed here.

It is noted that the above-mentioned lifetime measured in laboratories
is for neutral atoms.
Since $^{44}$Ti is a pure orbital-electron-capture decay isotope,
its lifetime depends on the electronic environment in the evolutional
course of a supernova remnant.
For example, a fully-ionized $^{44}$Ti is stable, 
and the lifetime of $^{44}$Ti in
the Hydrogen-like ionization state becomes longer by a factor of 2.25
than that of the neutral $^{44}$Ti
(see \citet{2004NewAR..48...69M}).
Let us briefly consider the effect of $^{44}$Ti ionization on our problem.

In young supernova remnants, the reverse shock propagates inward
through the ejecta and the resulting increase in temperature 
and density may lead to highly ionized ejecta material
through thermal collisions with free electrons.
A high-degree of ionization may then result in a longer lifetime of $^{44}$Ti,
which would significantly alter the inferred \Ti mass.
In fact, H-like and He-like Fe ions
have been observed in Cas A (see, e.g., \citealt{2004ApJ...615L.117H}).
Because the electron binding energies of Ti are smaller than
those of Fe, it is easier to ionize Ti than Fe.
Accordingly, \Ti  atoms in Cas A may be expected to be
in such high ionization states at least in part if they are
accompanied by the highly ionized Fe (this is expected
because \Ti is synthesized at the same location as where $^{56}$Ni
is also produced in the innermost region of a supernova).

Since the present-day radioactivity was entirely affected by the history
of various ionization stages and their duration time for which the $^{44}$Ti 
has experienced through the evolution, detailed discussion requires
numerical simulations as was done by 
\citet{1999A&A...346..831M} and \citet{2001NuPhA.688...58M}. 
However, we can get a rough idea of the ionization effect on the
radioactivity by using the result of simple linear analysis,  i.e.,
eq.~(7) of \citet{2004NewAR..48...69M}:
\begin{equation}
\Delta F_{\gamma}/F_{\gamma} =  (t/\tau - 1) \Delta \tau/\tau, 
\label{eq:dflux}
\end{equation}
where we have replaced the radioactivity $A$ and the decay rate $\lambda$
appeared in eq.~(7) of \citet{2004NewAR..48...69M} with the
$\gamma$-ray flux $F_{\gamma}$ and the $^{44}$Ti lifetime $\tau$,
respectively.
In eq.(~\ref{eq:dflux}), 
$t$ is the age of a SNR, $\Delta F_{\gamma}$ is the change of
the flux by ionization, $\Delta \tau$ is that of the lifetime.

Note that $\Delta \tau$ is always positive because the ionization always
increases its lifetime.  As was pointed out by the above authors,
the sign of $\Delta F_{\gamma}$ is then determined by that of
the term in the parenthesis in the right-hand side of
eq.(~\ref{eq:dflux}).
This means that the flux is {\em enhanced} by the ionization
when a SNR is {\em older} than the $^{44}$Ti lifetime, and that
the flux is {\em reduced} when it is {\em younger}.

Our concern here is whether the effect of ionization on the lifetime
of $^{44}$Ti in SNRs can reduce the disagreement between the
observed $^{44}$Ti Galactic map and the model's map or not.
From the above arguments, we can easily understand that the
discrepancy may be diminished if the $\gamma$-line fluxes in
Fig.~\ref{fig:Expected_ti44map} could be smaller 
which may be realized if most of 
the $\gamma$-ray detected SNRs in Fig.~\ref{fig:Expected_ti44map}
are younger than the $^{44}$Ti lifetime.

To get a rough idea, we performed a calculation
in which all parameters are the same as employed for
Fig.~\ref{fig:Expected_ti44map} except
1) the fluxes are multiplied by a factor 0.5 for SNRs with ages less
than 100 y,
and 2) the fluxes are multiplied by a factor 2 for SNRs with ages between
200 and 400 y.  The selection of 200-400 years old SNRs as enhanced
targets here is because the effect of the ionization due to the reverse
shock is
considered to be distinguished for these ages and the further inclusion of
the enhanced-flux effect on SNRs older than 400 y only makes the discrepancy
larger (see \citet{1999A&A...346..831M} for details).
Simulating a $10^5$ galaxy sky, we found that a $\gamma$-ray detector with
a detection limit of $1 \times 10^{-5}$ ph cm$^{-2}$ s$^{-1}$
would have a probability of detecting
0, 1, 2 $^{44}$Ti sources of 0.0012, 0.008, and 0.026, respectively.
Therefore, from this simple analysis it is suggested that the disagreement
cannot be compensated by the ionization effect; in effect it becomes
worse in our simple calculations above than 
the reference calculations of Fig.~\ref{fig:Expected_ti44map}.

A more precise estimate requires the knowledge of the temperature and
the density evolution of a supernova remnant, and the distribution of
$^{44}$Ti in it.  However, any detailed calculations taking into account
the retardation of $^{44}$Ti decay 
due to ionization will not alter the situation
better: In any case, the older SNRs whose fluxes may be enhanced always
dominate in number the younger SNRs whose fluxes may be decreased.

\subsubsection{Nucleosynthesis reaction rates}
\label{sec:reactionrates}

Estimates of yields of $^{44}$Ti from nucleosynthesis in supernovae depend
crucially on key nuclear reaction rates, and uncertainties in these rates limit
our ability to constrain the supernova rate.
\citet{1998ApJ...504..500T}
studied the sensitivity of $^{44}$Ti yields in alpha-rich freezeouts to
uncertainties in nuclear reaction rates.  They did this by computing the
alpha-rich freezeout with reference values for the reaction rates and then
comparing these results with ones from calculations with individual 
rates varied
upwards and downwards by a factor of 100 from their reference values.
The results were that the production of $^{44}$Ti
was most sensitive to the rates for the following reactions:
$^{44}$Ti$(\alpha,
p)^{47}$V, $\alpha(2\alpha,\gamma)^{12}$C,
$^{44}$Ti$(\alpha,\gamma)^{48}$Cr,
and $^{45}$V$(p,\gamma)^{46}$Cr for matter with equal numbers of neutrons and
protons ($\eta = 0$).  For neutron excess $\eta$ greater than zero,
the importance of the reaction $^{45}$V$(p,\gamma)^{46}$Cr drops, but other
reactions become more important.  In particular, these reactions are
$^{12}$C$(\alpha,\gamma)^{16}$O, $^{40}$Ca$(\alpha,\gamma)^{44}$Ti,
$^{27}$Al$(\alpha,n)^{30}$P, and $^{30}$Si$(\alpha,n)^{33}$S.

For our purposes, the relevant question is how much the $^{44}$Ti may vary
from current supernova models given these uncertainties.  Motivated by the
work of \citet{1998ApJ...504..500T}, \citet{2000PhRvL..84.1651S} measured the
cross section for the $^{44}$Ti$(\alpha,p)^{47}$V reaction at the
astrophysically relevant energies.  They found that the experimental cross
section for this reaction was a factor of two larger than in the rate
compilation of \cite{1987conf..ANA..525T} used 
in the \citet{1998ApJ...504..500T}
calculations.  From this result, \citet{2000PhRvL..84.1651S} inferred a
25\% reduction in the amount of $^{44}$Ti produced in alpha-rich freezeouts
in supernovae.  Other of the key reactions found by \citet{1998ApJ...504..500T}
had similar sensitivities of $^{44}$Ti yield to reaction rates; therefore,
if other experimental reaction rates are also a factor of $\sim 2$ different
from the theoretical calculations, we can expect similar $\sim 25\%$ effects
on the $^{44}$Ti yield.  From these results, we might thus conservatively
expect the $^{44}$Ti yield to be uncertain by less than a factor of $\sim 2$
due to reaction rate uncertainties.
Such a conclusion is supported by the study of the reaction rate sensitivity of
nucleosynthesis yields in core-collapse supernovae by
\citet{1999ApJ...521..735H}.
These authors compared the yields from core-collapse supernova models using
two different reaction rate libraries.  For the 15 M$_\odot$ stellar model
studied, the two calculations gave $^{44}$Ti yields that agreed to within
20\%, in spite of the fact that many individual nuclear reaction rates differed
by a factor of two or more between the two rate compilations.  

On the other hand, \citet{2005nucl.ex...9006N} have recently measured the
$^{40}$Ca($\alpha,\gamma)^{44}$Ti reaction cross section in the energy range
for nucleosynthesis in supernovae.  In that energy range, the authors find
that the reaction rate is 5 - 10 times larger than the previously
used theoretical rate calculated from a statistical model
\citep{2000NuPhA.675..695R}.  This large difference between
the experimental rate and the theoretical rate may be due to the fact that the
low level density in the $^{44}$Ti compound nucleus limits the applicability
of the statistical model for theoretical predictions for the rate of this
reaction.  In any event, the larger rate increases the yield
of $^{44}$Ti by a factor of $\sim 2$ in the stellar models
\citet{2005nucl.ex...9006N} explored.  Such a large increase in the
$^{40}$Ca$(\alpha,\gamma)^{44}$Ti reaction rate may allow normal core-collapse
supernovae to account for the solar system's supply of $^{44}$Ca; however,
this result would worsen the discrepancy between the observed
Galactic $^{44}$Ti gamma-ray flux and our predictions.

\subsubsection{The supernova explosion model}
\label{sec:explosionmodel}

In core-collapse supernovae, $^{44}$Ti production occurs by the alpha-rich
freezeout near the mass cut.  The location of the mass cut in the star will
then certainly affect the $^{44}$Ti yield.
Also important is the question of whether the simple-minded
notion of a mass cut at a single radial shell in the star
even makes sense in more realistic models that account
for large-scale fluid motions behind the stalled supernova shock prior to
the explosion and for stellar rotation.   These more realistic models suggest
that the material ejected from near the mass cut will in fact be a mixture
of parcels that arose from both inside and outside the mass cut.  We can
certainly expect variations in the entropies of those parcels, which, in turn
will have attendant variations in the $^{44}$Ti yield (e.g,
\citealp{2005ApJ...623..325P}).

For our purposes, the important issue to consider is how much variation can
we expect in the $^{44}$Ti yield from differences in the mass cut and
multi-dimensional effects.  One-dimensional models suggest that typically
half or more
of the $^{44}$Ti produced during the explosion might fall back on the remnant
(e.g, \citealp{2005nucl.ex...9006N}).  Similar results are possible for
the multi-dimensional models.  It is therefore quite conceivable that
yields of $^{44}$Ti from supernovae of the same mass
might vary by factors of $\gtrsim 2$ simply due to variations in the location
of the mass cut or multi-dimensional effects.  Of course, if the supernova
forms a black hole with mass greater than $\sim 2$ solar masses, it will
swallow up its innermost material and, thus, most or all of the $^{44}$Ti (and
$^{56}$Ni) it produced.  Such supernovae would be dim in both visible and
gamma radiation.  Perhaps the Galactic $^{44}$Ti map is indicating that
supernovae over the last few hundred years have been predominantly of
this type.

\subsection{Supernova Homogeneity}
\label{sec:homogeneity}

Although often taken for granted, homogeneity among supernovae of a type
  remains an open issue:  
   For thermonuclear supernovae, {\it light curves} have been found to be 
   fairly similar \citep{1998ARA&A..36...17B}. 
   Their successful empirical relative adjustment 
   through a light-curve-decline parameter apparently makes them 
   ``standard candles'' over the full range of cosmic evolution
   (This is the basis for the determination of cosmic expansion history,
   see e.g. \citet{2004MNRAS.350..253D}).
   The homogeneity of the {\it r-process elemental abundance pattern}
   in low-metallicity stars suggests that the r-process, which is commonly
   attributed to core collapse supernovae, 
   also presents a fairly well-regulated
   nucleosynthesis environment \citep[e.g. ][]{2002ceii.conf..205T}.
   On the other hand, the $^{56}$Ni masses ejected in supernovae
   appear to scatter, within \about~30\% for SNIa \citep{2005ApJ...623.1011B}, 
   and for
   core collapse events over a wide range from 0.01 to 1 \Msol 
   \citep{1995ApJS..101..181W, 1995ApJ...448..315W, 1996ApJ...460..408T},
   suggesting 
   more variability in the core collapse nucleosynthesis than in 
   thermonuclear explosions 
   \citep{1997NuPhA.621..467N, 2004NewAR..48..605T}. Still,
   for supernovae of type Ia alternative model types are also discussed that 
   would produce quite distinctly different nucleosynthesis products than 
   central carbon ignition in Chandrasekhar-mass white dwarfs 
   \citep{1982ApJ...257..780N, 1995ApJ...452...62L, 1994ApJ...423..371W}.
   
   In summary: supernovae of both types are quite homogeneous 
   in some of their characteristics,
   but anomalies suggest a deeper study of physical regulations and their
   observational impact. 

\subsubsection{2-3D Effects in Core Collapse Supernovae}
\label{sec:asymetries}

 \citet{1997ApJ...486.1026N}
in their explosive nucleosynthesis
calculations of 2-D axisymmetric Type II supernova found that
materials engulfed by energetic shock waves along polar
directions undergo higher temperatures (or higher entropy per baryon
\citep{2000ApJ...541.1033F})
to produce a higher amount of \Ti
than in spherical explosions.
Recently, \citet{2003ApJ...598.1163M} also studied hydrodynamics
and explosive nucleosynthesis in bipolar supernova/hypernova explosions.
Their bipolar models produce a large amount
($>$10$^{-4}$ M$_{\sun}$) of \Ti
and at the same time eject a relatively small amount
($\sim$0.1-0.2 M$_{\sun}$) of $^{56}$Ni.
These features of 2D supernova models 
inspire \citet{2004inun.conf...15P} to suggest that
the ``missing \Ti" problem (to be in concordance with SN1987A and Cas A
observations and also to account for the $^{44}$Ca solar abundance)
could be solved by avoiding the overproduction of $^{56}$Ni.
More systematic studies are therefore required under
variety of progenitor masses, explosion energies, metallicities,
and other physical variables.
As this 2D effect could elude the discrepancy between the \Ti production in
spherical supernova models and the amount 
observed from Cas A SNR in gamma-line fluxes
and the amount deduced from  $^{44}$Ca in solar abundance,
the axisymmetric explosion also seems to be the natural consequences
of rotation and magnetic field effects during pre-supernova phase
\citep{1981A&A...103..358M, 1994ApJ...434..268Y}
and the neutrino-driven convection \citep{1995ApJ...450..830B}
for the core-collapse supernovae.
Furthermore, evidence to support the axisymmetric explosion 
have been inferred from the measurements of
pulsar velocities \citep{1996AIPC..366...25B, 2004cosp.meet.1152H}
and from the jet features observed in radio and X-ray images from 
several supernova remnants \citep{1998MNRAS.299..812G}.
With various degrees of Rayleigh-Taylor instabilities 
that may develop in core-collapse supernova
(either in 2-D or 3-D simulations),
we expect there would be
a wide distribution of \Ti production in supernova events
as inferred indirectly by the pulsar velocity distribution.
Thereby the task is to find \Ti distribution production
with supernovae synthesizing higher than typical amount of \Ti produced in 
a spherical model 
(to explain the amount of $^{44}$Ca in solar abundance and Cas A SNR),
but also produces small amount of \Ti for the most-recent supernova so their
gamma-line emissions are too weak to be detected.
For this, we have to wait until the explosive 3-D nucleosynthesis can
be performed within reasonable time.

\subsubsection{Rare \Ti rich events}
\label{sec:tirich}

 The apparent deficit of \Ti remnants from typical supernovae as the main
 source of $^{44}$Ca
 is explained if typical supernovae are not the main source. Some rare
 type of event with a proportionately higher \Ti yield could be the major
 source
 and not leave detectable remnants today
 \citep[e.g.,][]{1988AIPC..170...98W}. For example, these could be
 He-triggered sub-Chandrasekhar-mass
 thermonuclear supernovae \citep{1994ApJ...423..371W}. 
 If their
 recurrence time is several times the \Ti lifetime, the Poisson
 probability of having
 none detectable now can be large. \citet{1994ApJ...423..371W} find some
 models with up to one hundred times higher \Ti yields than the typical
 supernova values discussed above. These need only occur now every one to
 two thousand years to provide the necessary $^{44}$Ca synthesis. Such a
 scenario, while potentially invisible to gamma-ray astronomy, would
 imply inhomogeneities among the relative abundances of $^{44}$Ca and other
 isotopes
 (see section \ref{sec:HecapIa}). A galactic survey at 6.9~keV could test this
 scenario \citep{2001ApJ...563..185L}. Another $\alpha$-rich freezeout nucleus, $^{59}$Cu, decays to $^{59}$Ni
 whose half-life is 75,000 years. 
  The nearest remnants of these rare objects from the past 10$^{5}$ years
 could be easily detectable in the subsequent cobalt K$_{\alpha}$ x-rays.
 
   Mainstream SiC grains, however, argue against rare producers of large amounts
  of \Ti being responsible for most of Galactic $^{44}$Ca, as the 
  $^{44}$Ca/$^{40}$Ca ratio in mainstream SiC grains does not vary much from 
  grain to grain; here sample size is much larger than for X grains, so 
  a more representative sampling may be assumed.
  (see Appendix \ref{sec:PresolarGrains}).

  We perform a simple simulation to test the viability of the rare-event scenario. 
In this model, we take the model that produces Fig. \ref{fig:ExpDisk}
(where the amount of \Ti in supernovae are as produced by
the supernova models explained in Sec. \ref{sec:Tiexpectations})
but modify the 10\% of core-collapse supernovae to produce 20$\times$
the \Ti of the supernova models. 
Therefore on average the amount of \Ti in this model is about a factor 2.7
of the model shown in Fig. \ref{fig:ExpDisk} to account for 
the $^{44}$Ca solar abundance.
This model gives a better agreement with the observed \Ti sky
than the model used to produce Fig. \ref{fig:Expected_ti44map} where
all supernovae produce 3$\times$ the amount of \Ti of supernovae
models.
As shown in the caption of Fig. \ref{fig:Expected_ti44map},
the probability of this model detecting 0, 1, and 2 \Ti sources are
0.012, 0.053, and 0.12, respectively.
This results show that the model is more probable than the model
of Fig. \ref{fig:Expected_ti44map} in explaining why COMPTEL and
INTEGRAL only detect one \Ti source (less than expected).
For comparison, the supernovae recurrence rates of 
the peak of the probability curve of 
the 1.157 MeV $\gamma$-line fluxes 
that are consistent
with the COMPTEL observed fluxes (see Appendix \ref{sec:fluxdataspace}
and the dotted curve in Fig. \ref{fig:ExpDisk}) 
for the model of Fig. \ref{fig:ExpDisk}, this model, and 
the model of Fig. \ref{fig:Expected_ti44map}  
are 36 yrs (Fig. \ref{fig:ExpDisk}), 40 yrs, and 58 yrs,  respectively.
This result is encouraging that a set of supernovae models  with
a rare type supernova that produces 
most of the solar $^{44}$Ca   
gives a closer recurrence rate to the one implied by the
historical supernova record ($\sim$17 yrs)
than the model with 3$\times$ amount of \Ti of
the calculated supernova models. 
Still we have no good explanation for
why some small number of supernovae are so rich in \Ti.

\section{Conclusions}
\label{sec:conclusions}

The observed distribution of pointlike sources of \Ti in our Galaxy is
  inconsistent (at a probability near 10$^{-3}$) with the combined current understanding of \Ti production,
 specifically
  \begin{itemize}  
  \item core collapse supernovae eject $10^{-4}$ \Msol  of \Ti, 
  within a factor of two, and
  \item the galactic core collapse  SN rate is about 3 per century,
\end{itemize}
  assuming the last few centuries are representative of the steady-state \Ti
production. The inconsistency is exacerbated if we further demand, as
currently understood to be so, that the standard solar abundance of
$^{44}$Ca originates from \Ti-producing
supernovae. The larger discrepancy is because the product of the above two quantities falls short by a factor of 
three of the requisite current $^{44}$Ca production rate. The disagreement persists for any combination of rate and
yield whose product is the required value, unless the yield of
\Ti is very much higher and the rates very much lower, i.e., $^{44}$Ca is not made primarily by typical
supernovae events, but by very rare ones. These might include He-triggered detonations of
sub-Chandrasekhar SNe Ia, or rare variants of core collapses, perhaps those most departing
from spherical symmetry. Evidence of He-cap SNIa as source of the $^{44}$Ca abundance might be
identifiable in $^{44}$Ca/$^{40}$Ca ratios greater than solar in some mainstream SiC grains.

A future survey with gamma-ray line sensitivity of 10$^{-6}\ cm^{-2}\ s^{-1}$  
would be expected
to detect $\geq$10 sources (Fig. 2), 
and so could rule out $^{44}$Ca production by frequent supernovae
at confidence 5~10$^{-5}$. Regardless of specific assumptions, 
the Cas A supernova remnant as
the brightest \Ti remnant in the galaxy is a priori very unlikely. 
That the brightest SNR should be found in
the outer galaxy or that it is over 300 years old are each improbable. 
It suggests that yields higher
than suggested by many current calculations are possible, 
which, of course, makes the lack of other detectable
remnants even more puzzling.



\begin{acknowledgements}
  Part of this work was supported by 
  NASA Grant NAG5-6892, NAG5-10764, NAG5-13565 to Clemson University
  and DOE's Scientific Discovery through Advanced Computing Program 
  (grant DE-FC02-01ER41189).
  Research by DDC was supported by NASA's Origin of Solar Systems Program.
\end{acknowledgements}

\vfill\eject 



\appendix
\section{Simulations of \Ti Skies}
\label{sec:montecarlo}

\subsection{Monte Carlo Simulation Details}
\label{sec:simulations}

  In generating supernova events of our Monte Carlo simulations, 
we adopt the procedure shown by 
\citet{1987ApJ...317..710H, 1992ApJ...387..314M, 1993A&AS...97..219H}.
A random number between 0 and 1000 is uniformly generated to represent 
a supernova age  between zero and 1000~y ($\simeq$11.4$\times\tau_{Ti44}$)
where $\tau_{Ti44} = (87.7 \pm 1.7)$ yrs
\citep{1997NuPhA.621...92N, 1998PhRvL..80.2554G, 1998PhRvL..80.2550A}
is the mean life of $^{44}$Ti.
This range is large enough that a supernova age older than 1000~y does not
contribute to the Galactic $^{44}$Ti flux. 

Another random number is generated to choose the type of the supernova
event. Then, several random
numbers are generated to give us the location of the supernova according
to its spatial distribution.
A detail procedure in generating the locations of supernova
from a disk and spheroid populations of Type Ia events
can be found in the paper of
\citet{1987ApJ...317..710H, 1992ApJ...387..314M}.
The distance of the supernova location to the Sun 
can be calculated easily.

A random number is generated from a Gaussian distribution to give us
the peak bolometric magnitude of the supernova following the distribution
given in section \ref{sec:models}. 
The apparent magnitude then can be calculated 
knowing the location, the distance, extinction magnitude (from 
Hakkilla et al.'s empirical model), 
and the peak bolometric magnitude of the supernova.
We find in model {\bf A} (Section \ref{sec:models})
the fraction of SNe in the last millennium 
brighter than 
apparent magnitude 0 is $\sim$11\% 
(see Fig. \ref{fig:expdisk_3kpchole_HakkilaExt_mB_dist}).

\begin{figure*}[htp]
 \includegraphics[angle=90,width=0.45\textwidth]{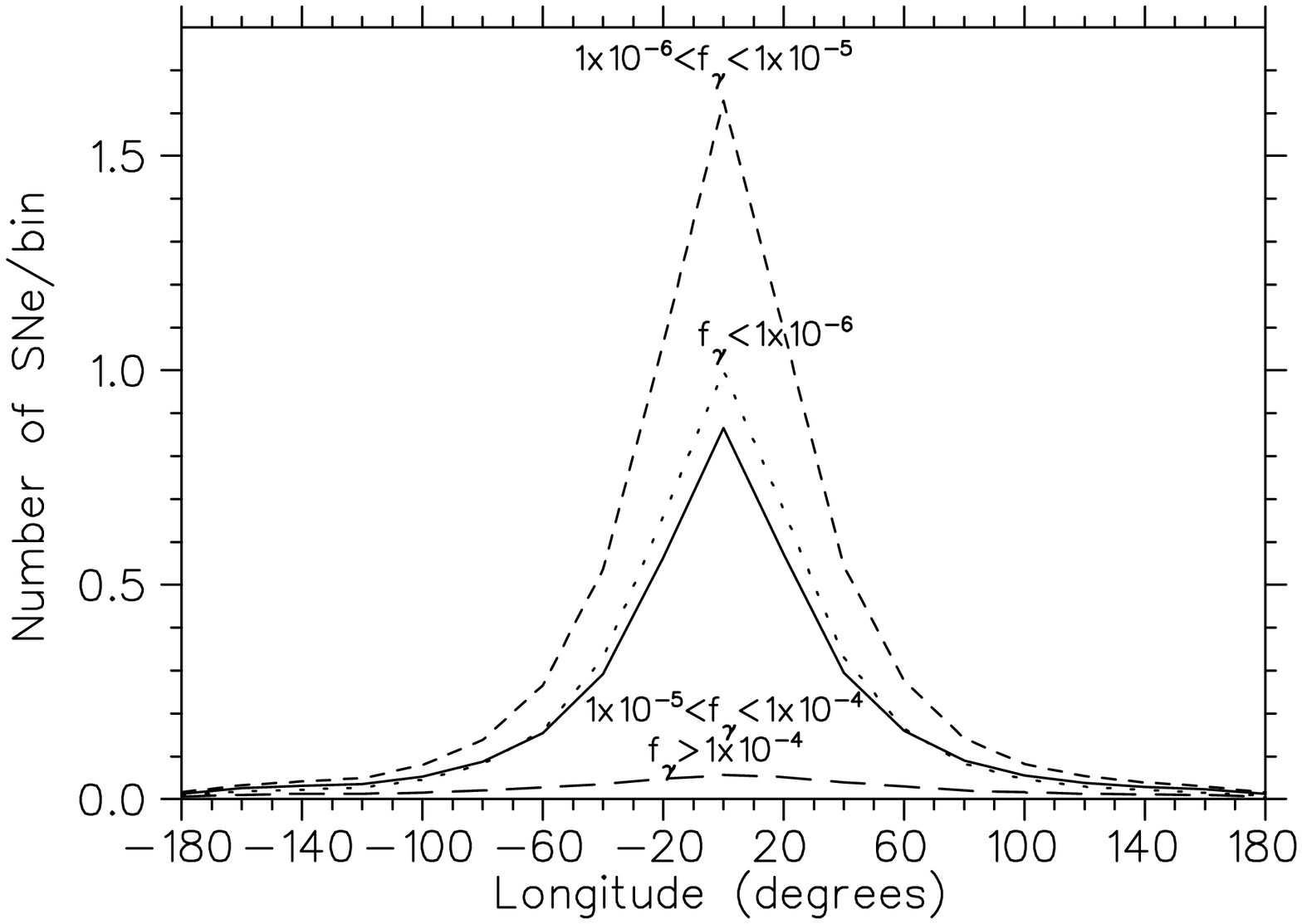}
 \includegraphics[angle=90,width=0.45\textwidth]{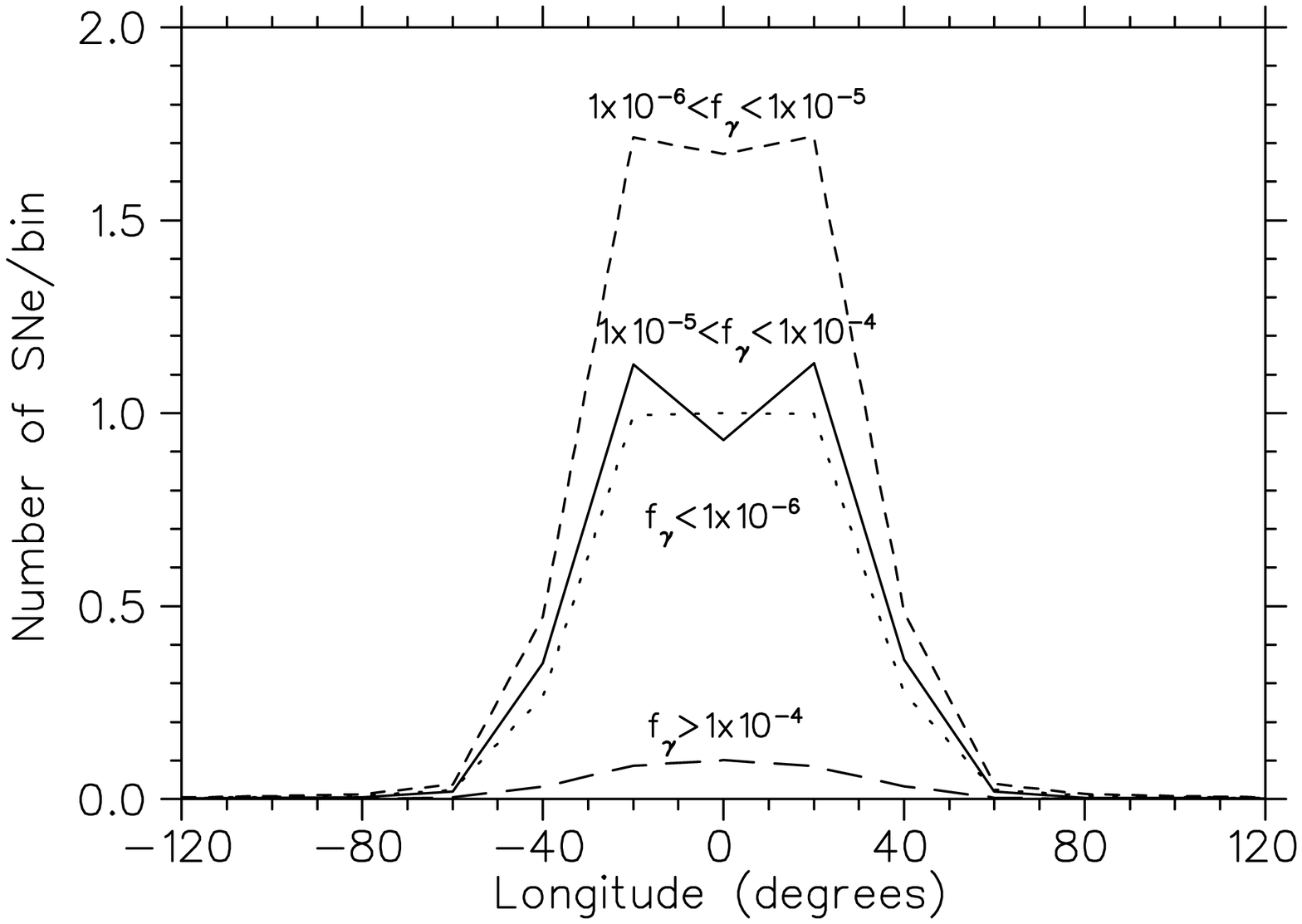}
  \caption{\it Same as Fig. \protect\ref{fig:longdist} but for models
  {\bf B} (left) and {\bf C} (right)}
  \label{fig:BClongdist}
\end{figure*}

The amount of $^{44}$Ti of the supernova is given by a uniform distribution
of $^{44}$Ti according to its type as described in section
\ref{sec:Tiexpectations}, 
{\it however note that the analysis in this appendix
does not use the multiplying factor of 3 that is used 
in \ref{sec:Tiexpectations}}.
$F_{\gamma}$, the $^{44}$Ti gamma-line flux at the Sun location,
then can be determined knowing 
the location, the distance, the age, and the amount of $^{44}$Ti of 
the supernova: \\
  $F_{\gamma} = 8.21\times10^{-3} \; M_4 \; exp(-t/87.7 yr)/d_{kpc}^2
  \; \gamma \; cm^{-2} \; s^{-1}$ \\ 
 where 
  $M_4$ is the $^{44}$Ti of the supernova in units of 10$^{-4}$ M$_{\sun}$
  and $d_{kpc}$ is the distance (in kpc) of the supernova from the Sun.
  We find that in Model {\bf A} the fraction of supernovae in the simulations
  with $^{44}$Ti $\gamma$-line fluxes larger than
  1$\times$10$^{-5}$ $\gamma \; cm^{-2} \; s^{-1}$ and
  3$\times$10$^{-5}$ $\gamma \; cm^{-2} \; s^{-1}$ are
  $\sim$11\% and $\sim$5\%, respectively (Fig. \ref{fig:CumulativeGflux}).
  Its longitude distribution along the Galactic plane in some relevant 
  $^{44}$Ti gamma-line flux bandwidths can be seen in
  Fig. \ref{fig:longdist}, while for models {\bf B} and {\bf C}
  (Section \ref{sec:models})
  are shown in Fig. \ref{fig:BClongdist}.

\begin{figure*}[htp]
\sidecaption
 \includegraphics[angle=90,width=0.7\textwidth]{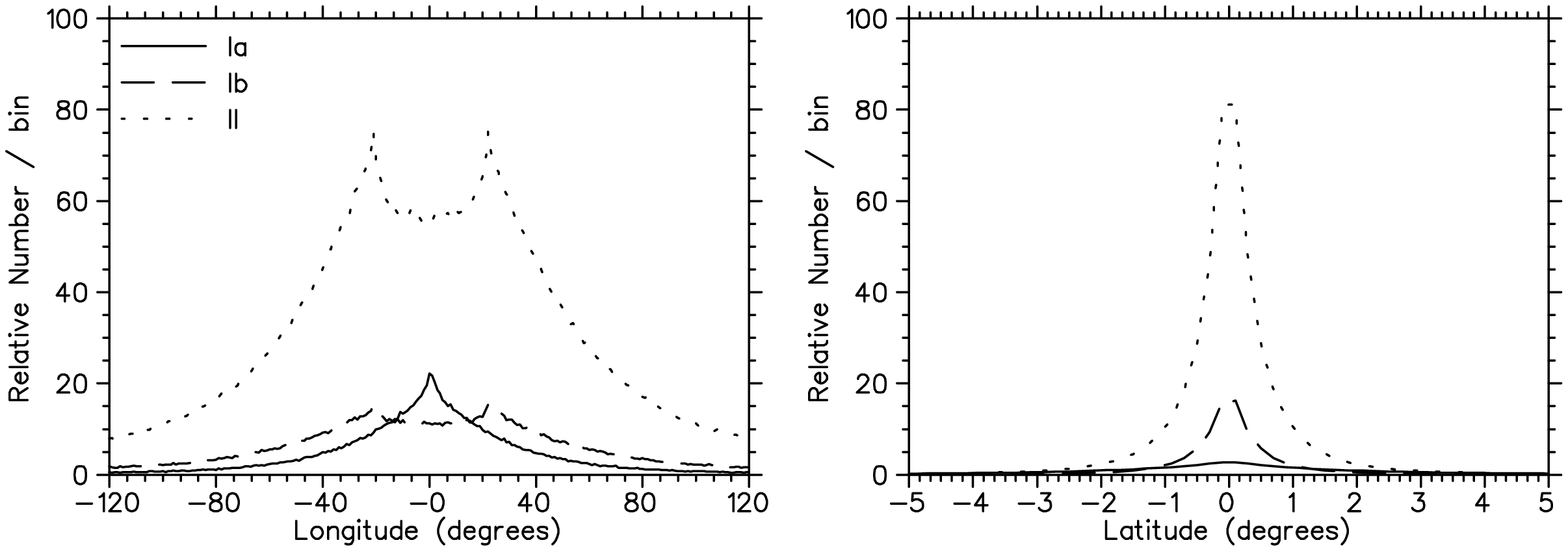}
   \caption{\it The distribution of supernovae in model {\bf A} in 
                Galactic longitude and latitude.} 
  \label{fig:expdisk_3kpchole_long_lat_dist}
\end{figure*}

We generate 1 million supernovae according to the above prescription and 
construct galaxies with numbers of supernovae between 1 and 300.
For each galaxy with a certain ${\it n}$ number of supernovae, 
the $^{44}$Ti gamma-line flux distribution $F(f_{\gamma},n)$
and the SN apparent magnitude distributed $M(m,n)$ can be extracted.

For a galaxy with SN recurrence time of T$_{rec}$ or 
average supernova rate, $\mu$ = 1000/T$_{rec}$ (we only simulate
supernova events in the last millennium as described above), 
the $^{44}$Ti gamma-line flux distribution, $F(f_{\gamma},\mu)$ 
can be obtained from 
$F(f_{\gamma},\mu)$ = $\sum_{n=0}^{\infty} \frac{e^{-\mu} \; \mu^n}{n!} \;
         F(f_{\gamma},n)$.

\subsection{Supernova Types and Spatial Distribution}
\label{sec:models}

Supernova events as the source of Galactic $^{44}$Ti can either
be Type Ia, Ib/c, or II SNe.
For this study, we choose a generally accepted (though uncertain) value
of the type ratio,
Ia : Ib/c : II = 0.1 : 0.15 : 0.75 
\citep{1994JRASC..88..369D, 1997MNRAS.290..360H}.
\citet{1996ApJ...462..266Y} and \citet{1995MNRAS.277..945T} 
using their chemical evolution model find 
that the ratio of the total number of Type Ia to Type II SNe of 0.12 gives
a good agreement with the observed solar abundance.
This is consistent with the observed Ia frequency which is as low as
10\% of the total SNe occurrence 
\citep{1991ARA&A..29..363V}. 
Monte Carlo representations of Type Ia SNe are generated using a nova
distribution template that traces the blue light distribution in M31
\citep{1987ApJ...317..710H}.
These populations form an axisymmetric
disc and a spherically symmetric bulge.
The spheroid density distribution follows:
\begin{itemize}
\item 1.25 $(\frac{R}{R_{\sun}})^{-6/8} \;
      e^{-10.093[(\frac{R}{R_{\sun}})^{1/4} - 1]}$ for $R \leq 0.03 R_{\sun}$
\item $(\frac{R}{R_{\sun}})^{-7/8} \;
      e^{-10.093[(\frac{R}{R_{\sun}})^{1/4} - 1]} \;
      [1 - \frac{0.08669}{(\frac{R}{R_{\sun}})^{1/4}}]$ for
       $R \geq 0.03 R_{\sun}$
\end{itemize}
and the disc density distribution follows:
\begin{itemize}
\item $n(\rho,z) \propto e^{-|z|/\sigma_z} \; e^{-(\rho-R_{\sun})/\rho_h}$
\item $\sigma_z$ = 0.325 kpc \citep{1981gask.book.....M} 
\item $\rho_h$ = 3.5 kpc \citep{1978AJ.....83.1163D}
\end{itemize}
The fraction of SNIa occurring in the spheroid is taken to be $\sim$1/7
of the total SNIa \citep{1980ApJS...44...73B}.

Supernovae of Types Ib \& II are associated with massive stars whose
birth places are exponentially distributed in height above the plane 
with a scale length of $\sim$100 pc.
Because the rate of core collapses is larger than the rate of
thermonuclear supernovae, we study several Type II distributions to ensure
that our results do not depend significantly on this choice.
We consider three cases:
\begin{enumerate}
\item Model {\bf A}: Exponential disk with no supernova within 3 kpc of the
      Galactic center, with radial scale length of 5 kpc 
      \citet{1997MNRAS.290..360H}:
\begin{description}
\item $n(\rho,z) \propto e^{-|z|/\sigma_z} \; e^{-\rho/\rho_d}$
\item $\sigma_z$ = 0.100 kpc 
\item $\rho_d$ = 5.0 kpc 
\end{description}
\item Model {\bf B}: Exponential disk with radial scale length of 3.5 kpc that
      produces an acceptable fit to the COMPTEL's $^{26}$Al
      $\gamma$-line map \citet{1995A&A...298..445D}.
\item Model {\bf C}: Gaussian-ring disk at radial distance of 3.7 kpc and
      radial distance scale length of 1.27 kpc \citep{1993ApJ...411..674T}.
\end{enumerate}

\begin{figure}[htp]
 \includegraphics[angle=90,width=0.45\textwidth]{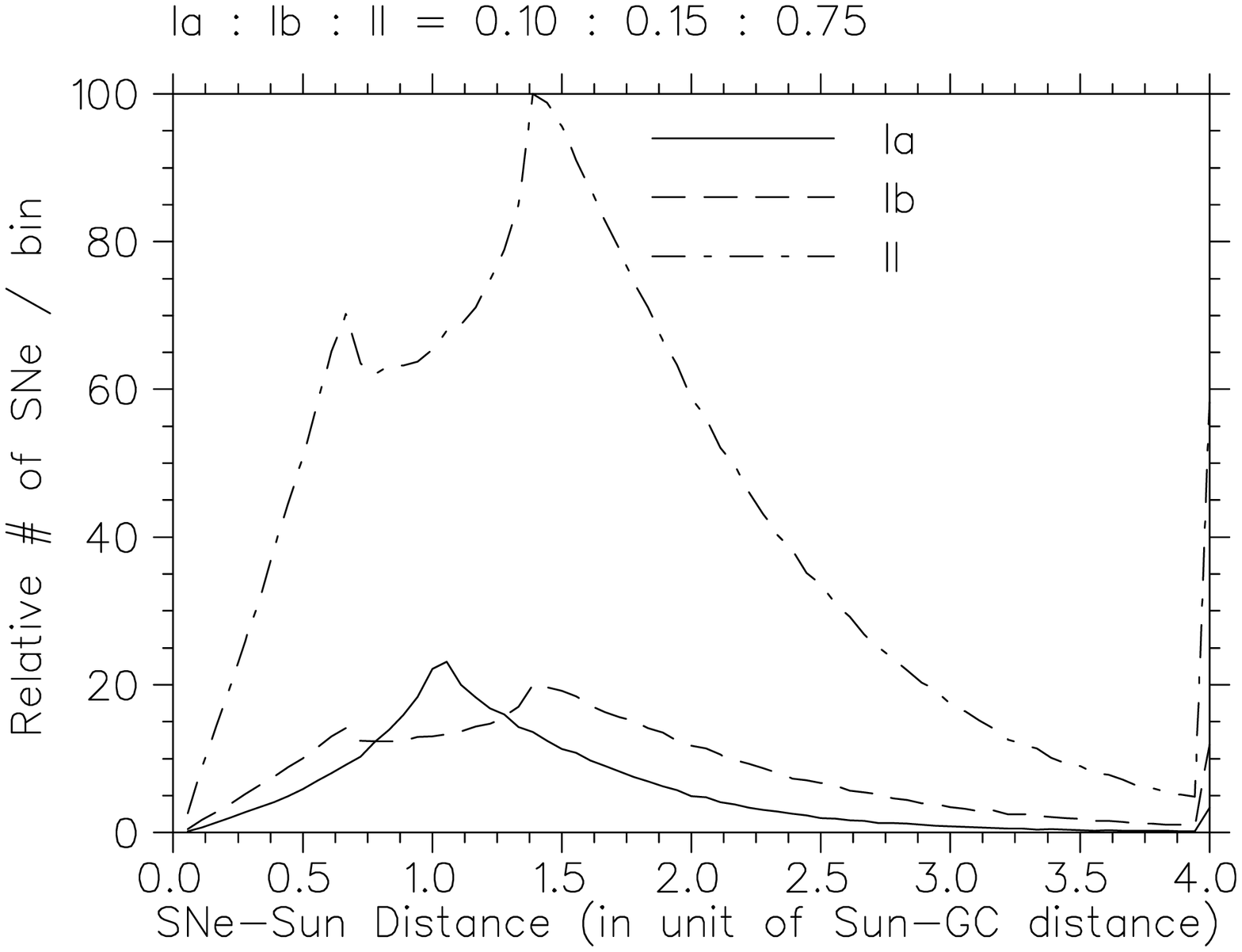}
   \caption{\it The distribution of supernovae distance in model {\bf A}.}
  \label{fig:expdisk_3kpchole_D_dist}
\end{figure}

The distribution of supernovae in model {\bf A} in 
directional is shown in Fig. \ref{fig:expdisk_3kpchole_long_lat_dist}
and in distance from the Sun is shown in
Fig. \ref{fig:expdisk_3kpchole_D_dist}.

\section{Comparison of Supernova Rates from \Ti Gamma-Rays and 
the Historical Record}
\label{sec:SNratecomp}
		 
In order to produce a consistent picture of the recent Galactic 
   supernova rate with the best known $^{44}$Ti supernova yields,
supernovae types and spatial distribution, and Galactic extinction model
we utilize our Monte Carlo simulation method to compare two observables
of the same phenomenon
\citep{1987ApJ...317..710H, 1992ApJ...387..314M, 1993A&AS...97..219H,
1999asra.conf...77T}.
We use the COMPTEL gamma-ray maps and 
the six historical Galactic supernova record of the last millennium
to constrain the range of Galactic supernova rates.

Similar to the above determination of \Ti flux distribution from 
a simulated set of supernova
events, other distribution such as the apparent magnitude distribution 
of supernovae
for a galaxy with an average supernova rate  $\mu$ can be constructed
(Section \ref{sec:historical}).
From this distribution we can determine the probability of optical 
detection for a certain detection-limiting apparent magnitude, 
such as magnitude 0, for example.
In this way, we can use the same underlying assumptions about 
supernova characteristics,
and compare two different observables, i.e. \Ti gamma-ray versus optical
detections of the supernovae at the adopted rates. Varying the rates then
so that they conform to the observational constraints, we obtain a handle
on systematic differences in observed supernova rates through these two
observational windows.

For the \Ti gamma-ray observations, we utilize the COMPTEL survey discussed
above (Sections \ref{sec:Tiproduction} and \ref{sec:gammaraydata}).
For the optical observations, we utilize the historical record summarized
in Table \ref{histrec}.

   \begin{table}
      \caption[]{Recent Galactic Supernova Record}
        \label{histrec}
      \[
         \begin{array}{p{0.3\linewidth}lcccc}
            \hline
            \noalign{\smallskip}
            Name            & Year  & Distance & l & b & Type \\
                            &       & (kpc)    &   &   &      \\
            \noalign{\smallskip}
            \hline
            \noalign{\smallskip}
            Lupus (SN 1006) & 1006  & 2.2  & 327.57 & 14.57 & \rm{Ia} \\
            Crab            & 1054  & 2.0  & 184.55 & -5.79 & \rm{II} \\
            3C58 (SN 1181)  & 1181  & 2.6  & 130.73 &  3.07 & \rm{II} \\
            Tycho           & 1572  & 2.4  & 120.09 &  1.42 & \rm{Ia}^a \\
            Kepler          & 1604  & 4.2  &   4.53 &  6.82 & \rm{Ib/II}^b \\
            Cas A           & 1680  & 2.92 & 111.73 & -2.13 & \rm{Ib}  \\
            \noalign{\smallskip}
            \hline
         \end{array}
      \]
       {\it $^a$: Studies of X-ray emission from Tycho's shocked ejecta
                  show the SNR was created by a Type Ia SN.
                 \citep{1998ApJ...497..833H, 2003ApJ...593..358B,
                  2005astro.ph.11140B}} \\
       {\it $^b$: The uncertainty of Kepler's SN Type was reviewed by
              \citet{2005ASPC..342..416B}.}
   \end{table}

  For comparison of the COMPTEL gamma-ray map with the results of Monte 
Carlo simulations, due to our limited computational power, for the
purpose to reduce systematic effects from regions of low exposure, and 
due to our limited ability to analyze the map, we perform two analysis. 
In the flux dataspace (Section \ref{sec:fluxdataspace})  we
use the flux information exclusively, without using the observed location 
of the gamma-ray source. 
In the map dataspace analysis (Section \ref{sec:Mapdataspace}, we only use
the COMPTEL map of inner galaxy ($|l|\leq$90$\degre$, $|b|\leq$30$\degre$)
which does not include the locations of the detected $^{44}$Ti $\gamma$-line 
from Cas A and GRO J0852-4642.
In order to see how consistent the model with the Galactic supernova
record, we perform the historical record analysis below.

\subsection{Map Dataspace Analysis}
\label{sec:Mapdataspace}
In map (imaging) analysis,
a simulated COMPTEL data set is produced by convolving
the directions and the $^{44}$Ti $\gamma$-line fluxes of 
the Monte-Carlo-generated
supernovae through the COMPTEL detector response.
The probability of the consistency of the simulated data with
the measured counts is carried
out by calculating the likelihood of the simulated data plus the
best-fit background model, and comparing this to the
likelihood of the background
model only \citep{1992daia.conf..241D}. 
The change in the likelihood gives the relative
probability that the specific realization of the model
is consistent with the measured COMPTEL data.
This process is repeated for at least 10$^4$ galaxies to obtain
the average relative probability of the model at a particular
supernova rate as dashed lines in Fig. \ref{fig:ExpDisk}, 
and \ref{fig:GaussianTC}.

\subsection{Flux Dataspace Analysis}
\label{sec:fluxdataspace}
 In flux dataspace analysis, in order to include the detected sources,
the locations of the $^{44}$Ti supernovae are not used,
flux information is used exclusively. The flux dataspace consists of
the two $^{44}$Ti detected fluxes from Cas A SNR and GRO J0852-4642
\citet{1998Natur.396..142I}
with other fluxes assumed to be zero (with which they are consistent.)
For each Monte Carlo Galaxy, a $\chi^2$ value is calculated by comparing
the two strongest fluxes with the fluxes measured from Cas A and
GRO J0852-4642 and the other fluxes are compared with null fluxes.
The probability of the model to be consistent with the COMPTEL fluxes
is determined from the $\chi^2$ and
using the number of degrees of freedom as the number of
independent 10$\degre\times$10$\degre$ image elements (fields of view)
of the supernova positions in the model.
This size of independent fields of view is obtained by calibrating
the size to produce the results of the
analysis of COMPTEL's map of the inner Galaxy.
Averaging the probabilities of 10$^6$ Monte Carlo galaxies, we obtain
the average probability shown in Fig. 
\ref{fig:ExpDisk}
and \ref{fig:GaussianTC}
as dotted lines.


\subsection{Historical Record Analysis}
\label{sec:historical}

\begin{figure}[htp]
 \includegraphics[angle=90,width=0.45\textwidth]{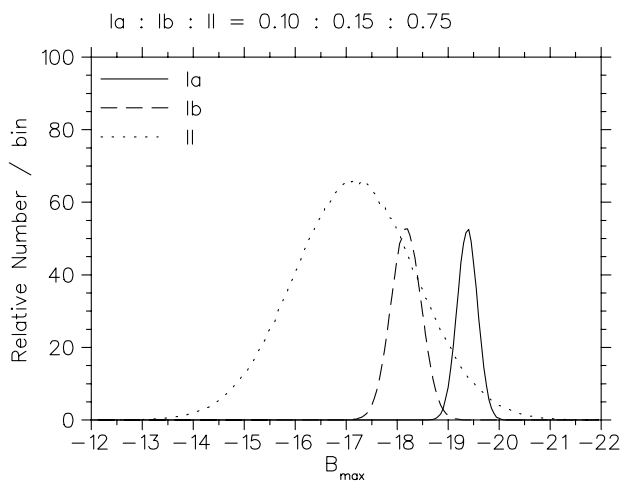}
   \caption{\it The distribution of the peak absolute magnitudes of 
                supernova in the B-band of model {\bf A}.}
  \label{fig:expdisk_3kpchole_Bmax_dist}
\end{figure}

The historical supernova record covering the last millennium shows a
total of only six Galactic SNe during that era (Table \ref{histrec}). Of course,
this small number is due to significant losses from extinction and
incomplete monitoring of the sky, especially during the early centuries.
In fitting this data,
we count the fraction of galaxies that have six SNe brighter than
magnitude 0. In this approach, we assume that historical SNe were
detected if they were brighter than magnitude 0.

\begin{figure}[htp]
 \includegraphics[angle=90,width=0.45\textwidth]{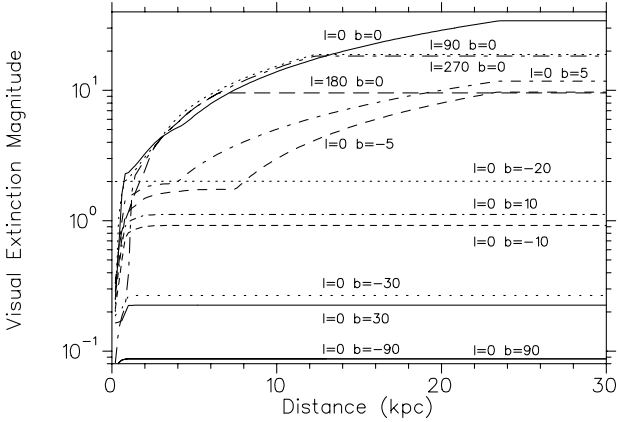}
   \caption{\it The Galactic visual extinction magnitude as a function of
distance from the Sun to various direction according to
empirical model of \citep{1997AJ....114.2043H}.}
  \label{fig:HakkilaExt}
\end{figure}

\begin{figure}[htp]
 \includegraphics[angle=90,width=0.5\textwidth]{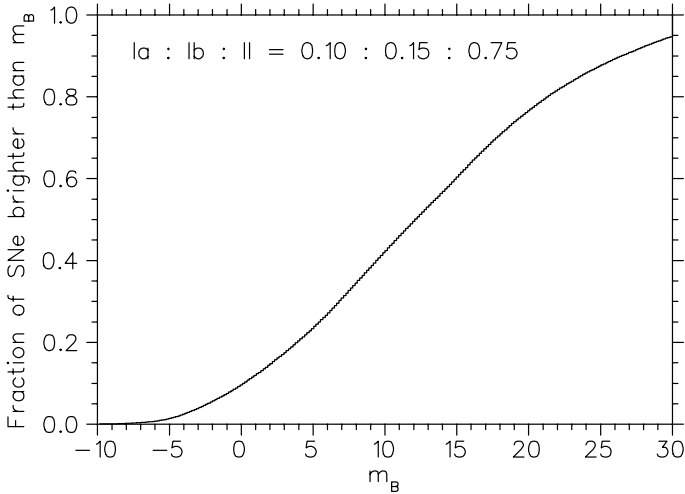}
   \caption{\it The apparent magnitude of supernovae in model {\bf A}
according to empirical extinction model of \citet{1997AJ....114.2043H}.}
  \label{fig:expdisk_3kpchole_HakkilaExt_mB_dist}
\end{figure}

  The peak absolute magnitudes of supernovae in the B-band are
approximated by Gaussian distributions.
For the mean values and the one standard deviation
we adopt for type Ia $M_B$ = -19.4, $\sigma$=0.2 
  \citep{1998ARA&A..36...17B},
for type Ib $M_B$ = -18.2, $\sigma$=0.3 
\citep{1994JRASC..88..369D},  and
for type II $M_B$ = -17.2, $\sigma$=1.2 
\citep{1991ARA&A..29..363V}.
Their distribution is shown in Fig. 
\ref{fig:expdisk_3kpchole_Bmax_dist}
Observed magnitudes of simulated SNe are obtained by convolving
absolute magnitudes through a Galactic extinction model of
\citep{1997AJ....114.2043H}.
In this extinction model, the total visual extinction
from the Sun's location to the Galactic Center is 11.64 magnitude
and the total visual extinction perpendicular to the Galactic plane at
Sun's location is $\sim$0.1 magnitude 
(Fig. \ref{fig:HakkilaExt}).
The cumulative distribution of supernovae having blue peak apparent magnitude
brighter than $m_B$ in model {\bf A}
is shown in Fig. \ref{fig:expdisk_3kpchole_HakkilaExt_mB_dist}.


The average fraction of the model that is consistent with
six events brighter than magnitude 0 is shown in
Fig. \ref{fig:ExpDisk}, 
and \ref{fig:GaussianTC} as solid lines.


\subsection{The Supernova Rates: Optical versus Gamma-Ray Constraints}
\label{sec:SNrates}

\begin{figure}[htp]
 \includegraphics[width=0.45\textwidth]{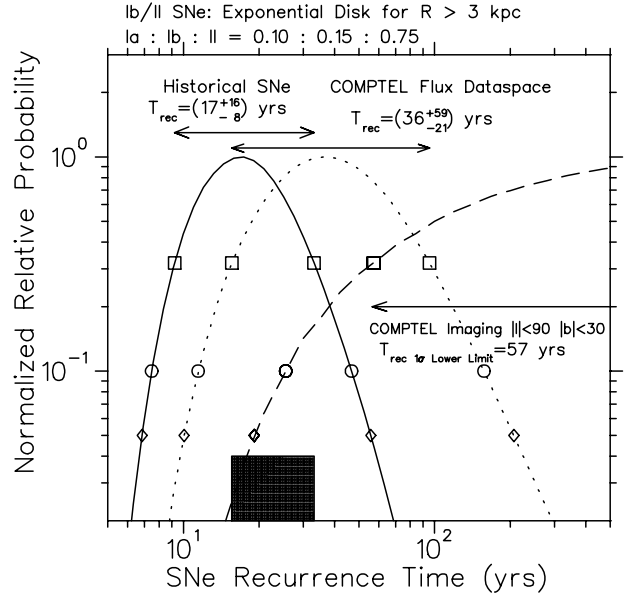}
 \caption{\it Normalized relative probability of Galactic SNe rate
  calculated from three different analysis vs. SNe recurrence time
  for model {\bf A}.
  The solid line is the probability of the model in which the
  number of SNe with $m_V \leq$ 0 is 6 in a 1000 year duration.
  The dashed-line is the probability inferred by maximum likelihood
  analysis of COMPTEL's 1.157 MeV $\gamma$-line inner galactic
  ($l\leq$90, $b\leq$30) image map.
  The dotted-line is the relative probability from chi-square test
  that the 1.157 MeV $\gamma$-line fluxes in the model are
  consistent with the COMPTEL observed fluxes.
  The labels of the curves show the optimal, 1$\sigma$ ranges, or
  the 1$\sigma$ lower limits of the SNe recurrence time.
  The boxes, circles, and diamonds on the curves show
  the 1$\sigma$, 90\%, and 95\% relative probabilities.
  The shaded area shows the range of the Galactic SNe recurrence
  time that is consistent with better than 1$\sigma$ probability 
  with each of the historical
  record and the COMPTEL flux dataspace analysis.}
  \label{fig:ExpDisk}
\end{figure}

Our  Galactic supernova record analysis leads to a most probable
Galactic supernova recurrence time of
$\sim$17, $\sim$16, and $\sim$13~y
based on models A, B, and C, respectively.
This implies a larger rate than given by previous
investigations based on the historical record.
For example, \citet{1994JRASC..88..369D} estimated
$\sim$3 SNe per century as also obtained by 
\citet{1994ApJS...92..487T}.
The rate is smaller than ours because
\citet{1994JRASC..88..369D}
considered 7 observed SNe within the last 2000~yrs and
assumed that the historical record
is 80\% complete.
A higher SNe rate of 5 SNe per century was obtained by
\citet{1997MNRAS.290..360H}
who include a
population of ``ultradim'' SNe in addition to 4 observed Galactic SNe
having $V<0$ and 80\% completeness within the last millenium.

The map and flux methods of estimating supernova rates based on
$\gamma$-ray
data appear to be significantly different, but they are statistically
consistent with each other for a wide range of supernova rates.
The imaging analysis gives a smaller rate than the flux analysis,
because the data used in the map analysis do not include
the Cas A and GRO J0852-4642 $\gamma$-line detections.
However, the probability estimate from the flux analysis
is a somewhat coarser estimate because the size of independent FOVs
used in determining the probability is not known exactly.
Ignoring the spatial information begs the question as to why the
two brightest SN are in the outer Galaxy rather than in the inner
Galaxy where we expected them.

In the flux dataspace analysis, where we ignore the expected spatial
      distribution of $^{44}$Ti remnants and consider only the measured
      flux distribution, we find a most probable supernova rate that is
      more compatible with standard values (such as the rate inferred by
the historical record), even for standard $^{44}$Ti yields.
      However, the COMPTEL map data indicate a lower SN rate
      ($\sim$1 SN/36 yr) than that suggested by
      the historical record ($\sim$1 SN/17 yr).
However model {\bf A} which does not have a supernova event in 
the inner 3 kpc radius has better agreement between the map
analysis and the historical record analysis.
This suggests that a model with less concentrated supernova
near the Galactic center than the model used here could give
a better agreement than we obtained here. 
Also, an extinction model with smaller extinction 
toward the Galactic center than Hakkilla et al.'s model
could give  a better agreement than what we present here.

\begin{figure*}[htp]
\centering
 \includegraphics[width=0.45\textwidth]{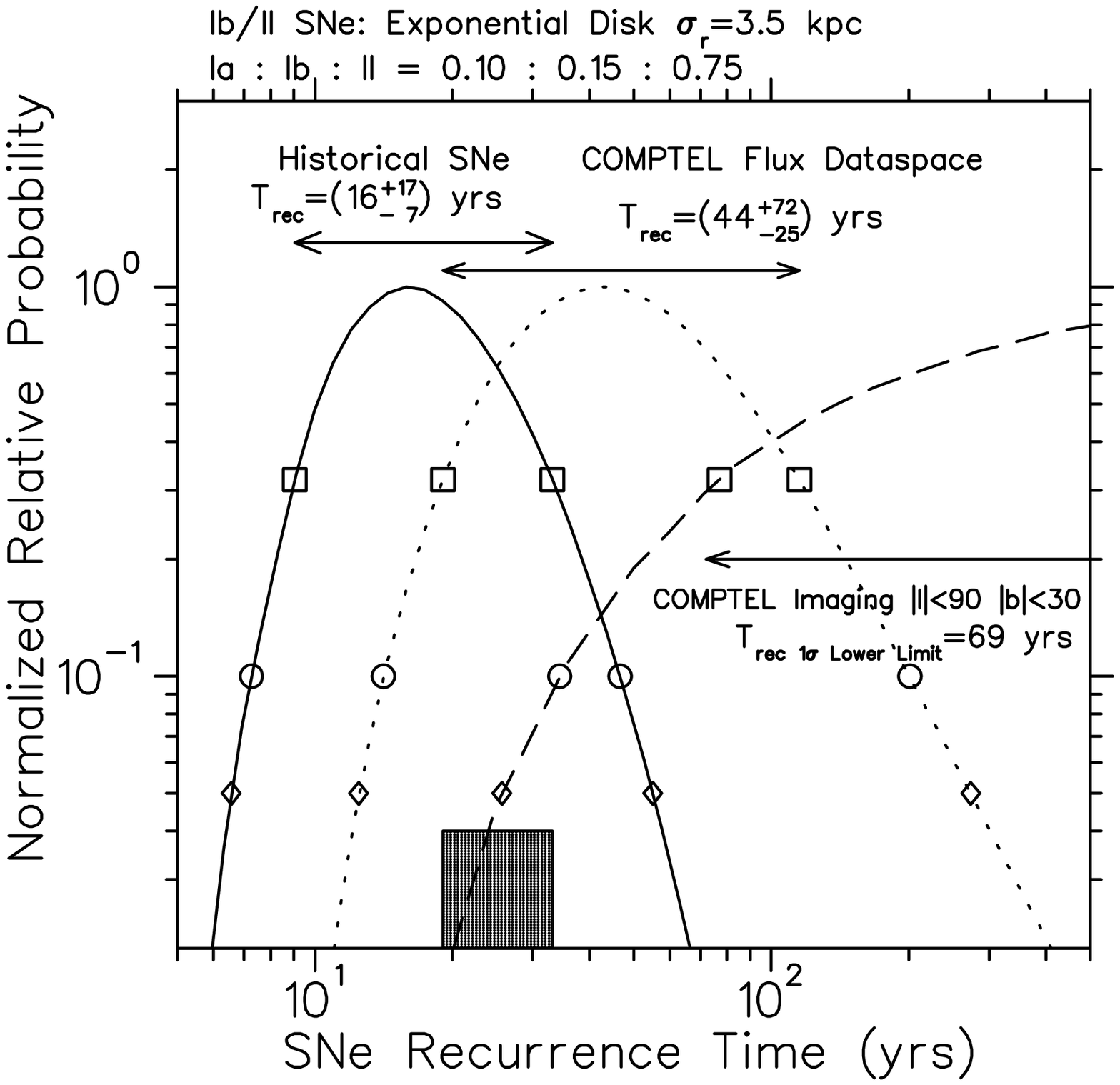}
 \includegraphics[width=0.45\textwidth]{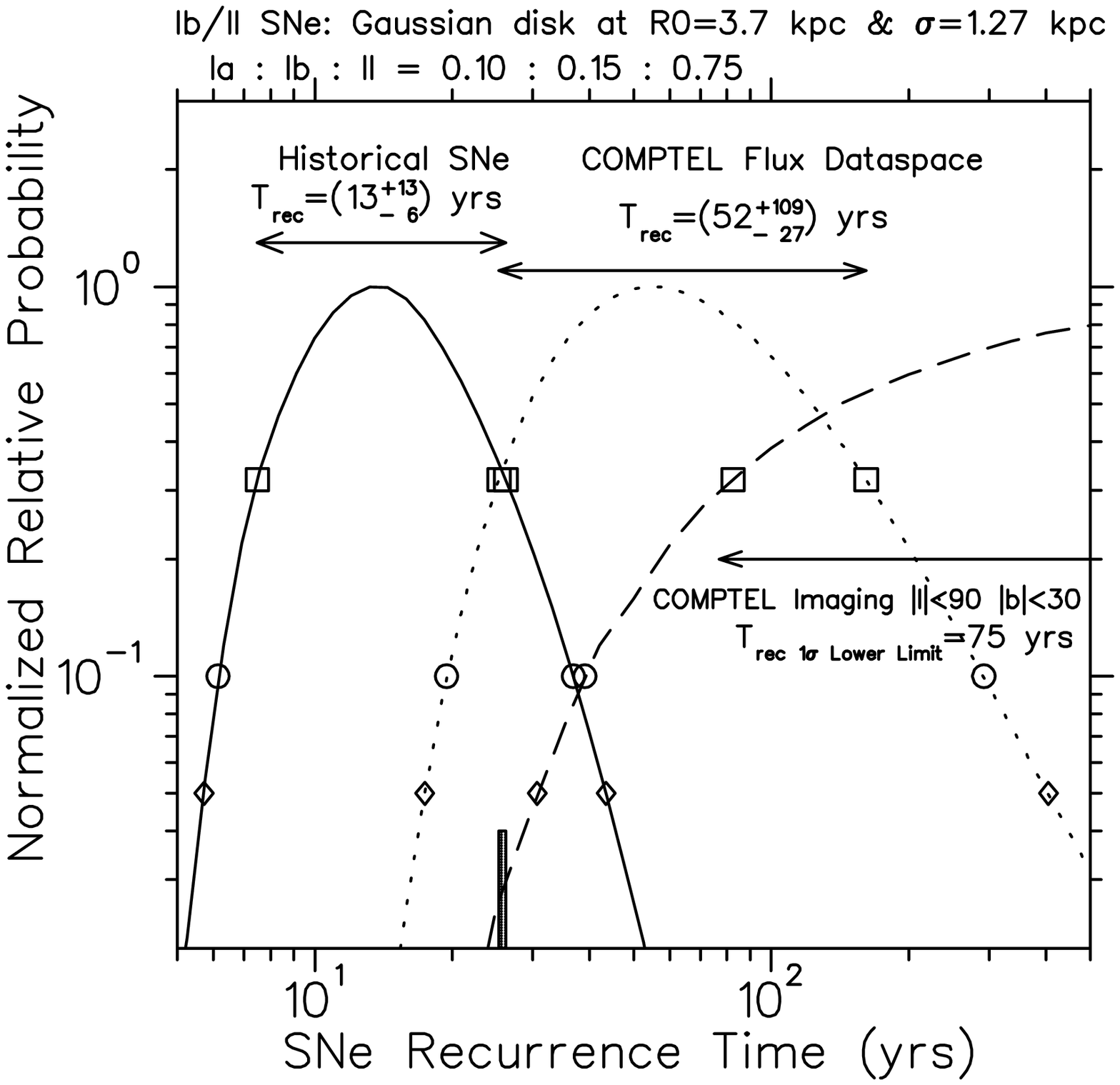}
  \caption{\it Same as Fig. \protect\ref{fig:ExpDisk} but for models
  {\bf B} (left) and {\bf C} (right)}
   \label{fig:GaussianTC}
\end{figure*}

 Based on chemical evolution studies,
      \citet{1996ApJ...464..332T}
      estimated that only
      $\sim$1/3 of the solar $^{44}$Ca abundance is accounted for.
      Models with a SN rate of $\sim$3 SNe per century
      and standard $^{44}$Ti yields
      fail to produce the solar $^{44}$Ca abundance.
      This rate, when confronted with the gamma-ray data (dashed line in
      Fig. \ref{fig:ExpDisk}
      and \ref{fig:GaussianTC})
      is too large:
      the COMPTEL gamma-ray data worsen an already serious
      problem. \citet{1996ApJ...464..332T}
      suggest 3 possibilities:
\begin{enumerate}
\item Increase the $^{44}$Ti yields
      by a factor of $\sim$3.
\item Increase the supernova rate by a factor of $\sim$3.
\item There is another source of $^{44}$Ca in the Galaxy.
\end{enumerate}
      Our analysis
      in Fig. \ref{fig:ExpDisk},
      and \ref{fig:GaussianTC}
      shows that
      the first and second option are not compatible with
      COMPTEL's $^{44}$Ti $\gamma$-line map, which
      would be brighter by a factor of 3 or exhibit a larger number
      of $^{44}$Ti hot spots than actually observed.
      We are thus left with the third option,
      to seriously entertain the idea that there
      exists some rare type of supernova
      (i.e., detonation of helium white dwarf),
      not realized in recent centuries, that produces very large amounts
      of $^{44}$Ti.

\section{Solar $^{44}$Ca constraints.}
\label{sec:Ca44constraint}
We now wish to estimate the effects of choices of chemical evolution model
parameters on the $^{44}$Ca constraint. 
We do this with analytic models \citep{1985ncnd.conf...65C}.
\citet{1994ApJ...424..200L}  used
that chemical evolution model 
to estimate the
current $^{44}$Ca production rate by requiring the yield that
gave exactly the solar abundance 4.55 Gyr ago. 
With model parameters k (infall contribution) and $\Delta$ (time delay) 
defining the shape of the galactic infall history, the current production is
\begin{equation}
p(^{44}Ca)= \, X_\odot(^{44}Ca) \, \frac{k+1}{\Delta} \, M_{g}(T_G) \ 
[ \frac{T_\odot+\Delta}{\Delta} -
(\frac{T_\odot+\Delta}{\Delta})^{-k}   ]^{-1},
\end{equation}
where T$_G$ is the age of the Galaxy, T$_\odot$=T$_G$-4.55~Gyr,
and M$_g$(T$_G$) is mass of interstellar gas participating in
star formation now. In this formulation, the $^{44}$Ca
production depends mainly on the 
infall parameter k= [f(t)/M$_{G}$(t)] (t + $\Delta$)  and the
gas mass, and weakly on $\Delta$ and the age of the Galaxy. 
The closed-box model, k=0, is ruled out by observations, and
k~=~2--4 is favored from a number of considerations 
\citep{1993ApJ...415L..25C}.
For the total $^{44}$Ca production rate, therefore, a low extreme value of 
p($^{44}$Ca)=1.1~10$^{-6}$\Msol yr$^{-1}$ is obtained for k=1,
$\Delta$=1.0~Gyr, M$_G(T_G)$=4~10$^{9}$\Msol, and
T$_{G}$=13~Gyr. 
An upper limit is estimated as
p($^{44}$Ca)=1.2~10$^{-5}$\Msol yr$^{-1}$ for  k=4,
$\Delta$=0.1~Gyr, M$_G(T_G)$=1~10$^{10}$\Msol, and
T$_{G}$=10~Gyr. 
Our favored value is 
p($^{44}$Ca)=5.5~10$^{-6}$\Msol yr$^{-1}$, derived for  k=2,
$\Delta$=0.1~Gyr, M$_G(T_G)$=1~10$^{10}$\Msol, and
T$_{G}$=12~Gyr. 
As in previous such studies 
\citep[e.g.][]{1994ApJ...424..200L}
we assume that this equals the $^{44}$Ti production rate. (However, we note
the statement by \citet{1996ApJ...464..332T} that
one-half of $^{44}$Ca is made directly as $^{44}$Ca rather than
as $^{44}$Ti.)
In such models, the solar $^{44}$Ca abundance thus could be provided by, for
example, 2.8 SNe per century
at the present epoch ejecting on average 2~10$^{-4}$\Msol
of $^{44}$Ti.

\section{Presolar grains}
\label{sec:PresolarGrains}
       Do the presolar SiC grains \citep{2004ARA&A..42...39C}
 that condensed within the expanding
supernova interior 
\citep{1997ails.conf..287A, 1997Ap&SS.251..355C}
shed light on the {\Ti} origin? Called "X grains", each
represents a sample of selected portions of a supernova interior, but "what
supernova" and "what portions of it" are unidentified questions for
each X grain. 
\citet{2003ApJ...594..312D} present a physical condensation argument 
to identify the region within the supernovae; but their argument awaits
digestion by the condensation-chemist community.
The X grains certainly contain evidence of having condensed with
abundant live \Ti in the form of very large $^{44}$Ca/$^{40}$Ca ratios
\citep{1996ApJ...462L..31N, 1997Ap&SS.251..355C, 2000M&PS...35.1157H}
that can be explained only through the action of live $^{44}$Ti. This
proves their supernova origin, and was a predicted signature of
supernova origin for presolar grains \citep{1975Natur.257...36C}
from the
beginning.  If each grain contains sufficient Ca and Ti the SIMS
analysis can yield the $^{44}$Ti/$^{48}$Ti abundance ratio at the time
of condensation (Fig. 2 of \citet{1997Ap&SS.251..355C}
and Figs. 8 and 9 of \citet{2000M&PS...35.1157H}).
Many X grains having
$^{44}$Ti/$^{48}$Ti = 0.1-0.6 are found.  The production ratio required
by the assumption that all $^{44}$Ca within solar abundances is the
product of decay of \Ti is P($^{44}$Ti)/P($^{48}$Ti) = 0.72; therefore,
supernova X grains are found with ratios up to the required bulk
production consistent with that assumption.  Despite this it must be
clearly understood that the measured production ratio
P($^{44}$Ti)/P($^{48}$Ti) within material in a specific grain bears no
simple relationship to the bulk production ratio in the supernova
within which that grain condensed.  This illustrates the problem of not
knowing precisely what supernova material the grain's isotopes
reflect. 
However, this problem seems capable of eventual solution
\citep{2003ApJ...594..312D}.

\subsection{SNII X grains}
\label{sec:IIxgrains}

        The SiC X grains probably originated in core-collapse SNII. 
Within them the pure alpha-rich freezeout 
\citep[][Section VII]{1973ApJS...26..231W},
which produces the bulk of \Ti ejected from SNII,
has production ratio near P($^{44}$Ti)/P($^{48}$Ti) = 1-2 
\citep[see Fig. 23 of ][ Section 3]{1973ApJS...26..231W, 1998ApJ...504..500T};
so no X grain is pure
alpha-rich freezeout material, although some of them must contain a
significant fraction of their Ti from the alpha-rich freezeout.  On the
other hand, the production ratio during normal silicon burning is very
much smaller, near P($^{44}$Ti)/P($^{48}$Ti) = 0.01
\citep[][ Fig. 19]{1973ApJS...26..231W}.
Therefore, X grains having $^{44}$Ti/$^{48}$Ti near
0.01 or less may contain no alpha-rich freezeout material at all, but
instead condense from Si-burning ejecta rich in $^{28}$Si.

        In their study of 99 SiC X grains from the Murchison meteorite,
\citet{2000M&PS...35.1157H}
found that 25 contained enough Ti and Ca for
isotopic analysis with the CAMECA IMS3f ion microprobe in Bern. Now
that the new nano-SIMS ion microprobe is functional, we may hope for
even more complete future surveys with its higher sensitivity. Of the
25 having enough Ti and Ca, 5 revealed large and easily resolved $^{44}$Ca
excesses, corresponding to P($^{44}$Ti)/P($^{48}$Ti) $>$ 0.01. Although
the chemistry of condensation of SiC in supernovae has not been solved,
initial studies \citep{2003ApJ...594..312D} of the
location of C and Si suggest that no more than 20\% of SiC X grains
should be expected to contain P($^{44}$Ti)/P($^{48}$Ti) ratios in
excess of those available during normal O and Si burning.  That is, a
majority of the X grains should contain no alpha-rich matter even if
each supernova ejects such matter. If this be taken as so, the data suggest that
the parent supernova population responsible for the 100s of X grains
that have been studied did eject alpha-rich freezeout material in
addition to its Si-burning matter. If ejection of alpha-rich freezeout
matter were instead rare, we expect that the number of SiC grains
having large P($^{44}$Ti)/P($^{48}$Ti) ratios would be significantly
less than 20\%.  
Without more detailed understanding of the origin of
the X grains it seems plausible to take their evidence to suggest that
all SNII eject alpha-rich-freezeout  matter.

        If the SiC X grains do originate in SNII, their frequent
$^{44}$Ca excesses demonstrate that either the X grains all originated
in a single nearly supernova that did eject \Ti or originated in many
presolar SNII that mostly ejected \Ti (although not necessarily
always).  The properties of the  "mainstream SiC grains" enable us to
argue against the possibility of a single supernova as origin of the X
grains.  The much more abundant mainstream grains, which are thought to
have originated in presolar carbon stars, have been argued on the basis
of their Si isotopes to have been the result of a great many C stars.
Their Si isotopic compositions represent approximately the initial Si
isotopic compositions from which the intermediate-mass stars formed,
and their correlation between excess $^{29}$Si and $^{30}$Si is
explained by galactic chemical evolution effects that have produced a
wide range of Si isotopic compositions in a large number of C stars
\citep{1996ApJ...472..723T, 2003ApJ...598..313C}.
Because the mainstream grains comprise about
10$^{-4}$ of all interstellar Si, much of which is quite old, and
because the condensed SiC from the C-star phase is but a small fraction
of all Si ejected from stars, it seems that the lifetime of SiC grains
in the ISM is not short. If this be so, the lifetime of X SiC grains is
also not short, suggesting a substantial number of contributing
supernovae to the matter gathered into any cloud in the ISM.

\subsection{SNIa X grains}
\label{sec:Iaxgrains}

        For the question of \Ti gamma-ray hotspots the issue arises
whether the SiC X grains may instead have arisen from exploding SNIa.
After all, the high expectation for seeing \Ti supernova remnants shown
in Figure 1 came from the assumption that Type II are the sources of
\Ti nucleosynthesis and therefore also of the solar $^{44}$Ca
abundance. The absence of such sources could be explained if rare Type
I events produced much larger yields than are expected from Type II. Do
the supernova grains offer any guidance? 
\citet{1997ApJ...486..824C} showed
that isotopically good fits to X grains might originate within He caps
on exploding C,O white dwarfs.  Huge production ratios, up to
P($^{44}$Ti)/P($^{48}$Ti)=100, occur in those He shell zones having
peak T$_9>$1 (see their Fig. 6). These are too great for existing
observations of X grains, suggesting that SNIa are not their sources.
However, \citet{1997ApJ...486..824C} showed that good fits to all isotopes
require that the He caps undergo considerable post-explosive mixing
prior to condensation of the X grains. Their Table 2 shows production
ratios as small as 0.01 in an average of zones 1-8 (the coolest eight
zones) and near 10 in an average over all sixteen zones. (The reader
must note that much $^{48}$Ti production is listed by them under
$^{48}$Cr, its radioactive progenitor.) The isotopic possibility must
therefore be addressed that the $^{44}$Ti-rich X grains are from SNIa
He caps rather than from SNII.
The discovery of even a single X-type SiC grain containing an 
initial ratio $^{44}$Ti/$^{48}$Ti=5 or greater would demonstrate that
the He-cap SNIa do exist and that they can condense SiC. 
This would amount to an existence proof for these rare $^{44}$Ti producers. 
But no such grain has yet been detected.

\citet{1997ApJ...486..824C} pointed out that if the SNIa are near the
Chandrasekhar mass, the He cap can be no more than 0.01~\Msol\, so that
less than 10$^{-5}$\Msol of \Ti is ejected.  Such events would be both
rare and dim in \Ti lines. However, a much larger He cap is involved if
the white dwarf is sub-Chandrasekhar (larger) and the detonation begins
in a massive He cap. Large amounts of \Ti are ejected from such models
\citep{1994ApJ...423..371W},
so that the rates of such events are limited by the
requirement that they not overproduce the galactic $^{44}$Ca
abundance.  If this occurs, the rarity of these events in time could
account for their absence in the COMPTEL data (see also the
Discussion). The relevance of the X grains derives from the physical
unlikelihood of condensation of X grains in such He caps.  Their
expansion is very fast, and the radiation environment intense, so that
even the formation of molecules seems unlikely until the density has
become too low for the growth of a 1-micrometer grain of SiC.  
\citet{1997ApJ...486..824C} 
note that even the possibility of grain condensation would
seem to require 3-D explosive modeling in order that some of the He cap
can remain at low velocity ("slow He"). This entire problem will
require more study before plausibility of SiC X grains from SNIa can be
admitted. We therefore conclude that the $^{44}$Ti-bearing X grains
have arisen in SNII.

\subsection{ISM Inhomogeneity of $^{44}$Ca from He-cap SNIa}
\label{sec:HecapIa}

  We turn now to non-supernova Stardust whose $^{44}$Ca ratios reflect
the initial compositions of their donor stars.
        The supernova X grains seem to support the assumptions that led
to the conflict between the COMPTEL map (Fig. \ref{fig:timap_IyudinRingberg}) 
and the expected
hotspots (Fig. \ref{fig:Expected_ti44map}); 
namely, that X grains condensed within SNII, that
the number of contributing SNII was large, and that most ejected
alpha-rich freezeout matter.  
However, X grains do not demonstrate
that the bulk ejecta of SNII contain sufficient mass of {\Ti}  to
account for the natural $^{44}$Ca abundance.  
To test for the presence
of rare large {\Ti} producers \citep{1994ApJ...423..371W}
responsible for
making good the shortfall from SNII we turn to the mainstream SiC
grains with the following original argument. 
Rare SNIa responsible for roughly 2/3 of galactic $^{44}$Ca
plausibly result in Galactic inhomogeneities in the interstellar
$^{44}$Ca/$^{40}$Ca ratio that should be larger than those seen by
astronomers for elemental-ratio variations attributed to inhomogeneous
incorporation of SNII ejecta.  
It must be slower to homogenize the ISM if SNIa occurring every 3000 yr
contribute 2/3 of $^{44}$Ca abundance than for homogenizing ISM from
SNII occurring every 30 yr and making but 1/3 of the $^{44}$Ca abundance.
ISM regions temporarily enriched in the rare SNIa ejecta may make stars
having larger $^{44}$Ca/$^{40}$Ca initial ratios.
The mainstream SiC grains reflect the interstellar composition from
which the carbon stars formed.
That is, factor of two or more
variations in  $^{44}$Ca/$^{40}$Ca ratio should be present in the
initial compositions of stars.

  It is this expectation that the mainstream grains may speak against. 
\citet{2000M&PS...35.1157H}
measured Ca isotopes in 28 mainstream SiC grains (see their
Fig. 8) and found their  $^{44}$Ca/$^{40}$Ca ratios to be
indistinguishable from solar despite large Si isotope variations in the
same stars demonstrating that the grains come from different AGB C
stars having distinct chemical evolution histories for their initial
compositions \citep{1996ApJ...472..723T}.
Each mainstream grain was
consistent with solar $^{44}$Ca/$^{40}$Ca ratio and their average was
enriched 3.8\% in $^{44}$Ca, as expected for modest s-process enrichments
of AGB atmospheres by the third dredge ups. Because of the small numbers
of Ca atoms in mainstream grains, however, individual grain ratios were
uncertain by at least 20\%. Even so, 20\% is a small inhomogeneity for
rare events producing 2/3 of the bulk  $^{44}$Ca. To improve this data
base we call for high precision measurements of the $^{44}$Ca/$^{40}$Ca
ratio in a sample of mainstream grains with the new nanoSIMS ion
microprobes. Such a survey could reveal more precisely any variations
in initial $^{44}$Ca/$^{40}$Ca ratios in intermediate-mass stars
expected if the \Ti synthesis has instead been the result of a few rare
events of very large \Ti yield. The present data speak against that.
        The new arguments that we have presented here can all be
improved dramatically in the next few years. Although their message
today is not statistically certain, their sense is to support the
original conflict that we have displayed in 
Fig.  \ref{fig:Expected_ti44map}.
Because the  $^{44}$Ca/$^{40}$Ca ratio in mainstream grains from 
many AGB stars is very near to the solar ratio, we point out that
it is no longer possible to entertain the idea that the sun may
itself contain an anomalous  $^{44}$Ca/$^{40}$Ca ratio.
One can not blame the inability of standard models of the chemical
evolution of the galaxy to produce enough $^{44}$Ca by speculating
that the sun is abnormally rich in its $^{44}$Ca content.

It should be noted that a thorough study of stardust isotopic
inhomogeneity within the context of inhomogeneous galactic chemical
evolution has been presented by 
\citet{2005ApJ...618..281N}.
He shows in particular
that correlations between Si and Ti isotopic compositions found in
stardust can not be accounted for by inhomogeneous GCE. He does not,
however, discuss the variations in $^{44}$Ca/$^{40}$Ca
that are generated by his
model owing to the He-cap SNIa. On the other hand, his calculations may
not be ideal for this problem because his model admixes into a homogeneous
ISM  the same fraction for each supernova's ejecta 
(a= 1.7$\times$10$^{-5}$) after Monte Carlo sampling 
to obtain N=70 supernovae. Nonetheless, that paper shows
the potential power of stardust to delimit GCE inhomogeneities in general.
This approach should be reconsidered carefully when a good data set for
initial $^{44}$Ca/$^{40}$Ca  in stardust is available.

Mainstream SiC grains are probably not the best grains with which 
to measure initial stellar $^{44}$Ca/$^{40}$Ca abundance ratios. 
The mainstream SiC grains contain very little Ca because 
Ca is much less favored chemically than is Ti within 
thermally condensing SiC grains. 
The hibonite Stardust grains described by \citet{2005LPI....36.2200N},
on the other hand, contain much more Ca because Ca is 
an essential ingredient of the hibonite crystal structure.
Therefore, accurate measurements of $^{44}$Ca/$^{40}$C 
in hibonite grains could reveal better information relevant 
to ISM inhomogeneity of the $^{44}$Ca/$^{40}$Ca ratio.

\end{document}